\documentclass[aps,prd,nofootinbib,floatfix,showpacs,preprintnumbers,twocolumn]{revtex4}
\usepackage{bm}
\usepackage{latexsym}
\usepackage{dcolumn}
\usepackage{amsfonts,amssymb,amsmath}
\usepackage{graphicx,epsfig}
\usepackage{psfrag}  
\usepackage{multirow}

\def\beq{\begin{equation}}
\def\eeq{\end{equation}}
\def\br{\begin{eqnarray}}
\def\er{\end{eqnarray}}
\def\benu{\begin{enumerate}}
\def\eenu{\end{enumerate}}
\def\bi {\begin{itemize}}
\def\ei {\end{itemize}}
\def\l{\left}
\def\r{\right}

\def\nn{\nonumber} 

\def\pa{{\partial}}

\def\f {\frac}

\def\b  {\beta}

\def\cA {\mathcal A}

\def\cE {\mathcal E}
\def\eq#1{{Eq.~(\ref{#1})}}

\def\CGEV#1{{{\cal G}^{#1}({\cal E},r)}}
\def\CGmEV#1{{{\cal G}_{m}}^{#1}({\cal E},r)}

\newcommand{\lPl}{\ell_{_{\rm Pl}}}
\newcommand{\SBH}{S_{_{\rm BH}}}
\newcommand{\SBW}{S_{_{\rm BW}}}
\newcommand{\AHo}{{\mathcal A}_{_{\rm H}}}
\newcommand{\rHo}{r_{_{\rm H}}}
\newcommand\PiIV{P^{(i)}_{4}(r)}
\newcommand\PiVI{P^{(i)}_{6}(r)}
\newcommand{\FAH}{{\cal F}({\cal A}_{\rm H})}
\newcommand{\FAHFD}{{\cal F}^{\rm (4D)}({\cal A}_{\rm H})}
\newcommand{\FAHSD}{{\cal F}^{\rm (6D)}({\cal A}_{\rm H})}
\newcommand{\FAHOz}{{\cal F}_{_{\rm (0)}}({r}_{\rm H})}
\newcommand{\FAHSz}{{\cal F}_{_{\rm (2)}}({r}_{\rm H})}
\newcommand{\GAH}{{\cal G}({\cal A}_{\rm H})}
\newcommand{\GAHOz}{{\cal G}_{_{\rm (0)}}({r}_{\rm H})}
\newcommand{\GAHSz}{{\cal G}_{_{\rm (2)}}({r}_{\rm H})}
\newcommand\CGE{{\cal G}({\cal E},r)}
\newcommand\CGmE{{\cal G}_{(m)}({\cal E},r)}

\reversemarginpar

\begin{document}
\preprint{Preprint number: arxiv:0710.2013 [gr-qc]}
\title{Sub-leading contributions to the black hole entropy in the 
brick wall approach}
\author{Sudipta Sarkar}\email[E-mail:~]{sudipta@iucaa.ernet.in}
\affiliation{IUCAA, Post Bag 4, Ganeshkhind, Pune 411~007, India.}
\author{S.~Shankaranarayanan}\email[E-mail:~]{shanki@aei.mpg.de}
\affiliation{Max-Planck-Institut f\"ur Gravitationphysik,
Am M\"uhlenberg 1, D-14476 Potsdam, Germany.}
\author{L.~Sriramkumar}\email[Email: ]{sriram@hri.res.in}
\affiliation{Harish-Chandra Research Institute, Chhatnag Road,
Jhunsi, Allahabad 211 019, India.}
\date{\today}
\begin{abstract}
The brick wall model is a semi-classical approach to understand the
microscopic origin of black hole entropy.  
In this approach, the black hole geometry is assumed to be a fixed 
classical background on which matter fields propagate, and the 
entropy of black holes supposedly arises due to the canonical 
entropy of matter fields outside the black hole event horizon, 
evaluated at the Hawking temperature.  
Apart from certain lower dimensional cases, the density of states 
of the matter fields around black holes cannot be evaluated exactly.  
As a result, often, in the brick wall model, the density of states 
and the resulting canonical entropy of the matter fields are evaluated 
at the leading order (in terms of~$\hbar$) in the WKB approximation.  
The success of the approach is reflected by the fact that the 
Bekenstein-Hawking area law---viz. that the entropy of black 
holes is equal to one-quarter the area of their event horizon, 
say, $\AHo$---has been recovered using this model in a variety 
of black hole spacetimes. 
In this work, we compute the canonical entropy of a quantum scalar 
field around static and spherically symmetric black holes through 
the brick wall approach at the higher orders (in fact, up to the 
sixth order in~$\hbar$) in the WKB approximation. 
We explicitly show that the brick wall model generally predicts 
corrections to the Bekenstein-Hawking entropy in all spacetime
dimensions. 
In four dimensions, we find that the corrections to the
Bekenstein-Hawking entropy are of the form $\l[\AHo^{n} \log \AHo\r]$, 
while, in six dimensions, the corrections behave as $\l[\AHo^m 
+ \AHo^{n}\, \log \AHo\r]$, where $(m, n) < 1$.  
We compare our results with the corrections to the Bekenstein-Hawking 
entropy that have been obtained through the other approaches in the 
literature, and discuss the implications.
\end{abstract}
\pacs{04.70.Dy, 04.62.+v}
\maketitle

%%%%%%%%%%%%%%%%%%%%%%%%%%%%%%%%%%%%%%%%%%%%%%%%%%%%%%%%%%%%%%%%%%%%%%%%%%%%%%%

\section{Introduction and motivation}

The concept of black hole entropy was originally introduced by
Bekenstein~\cite{Bekenstein:1972c,Bekenstein:1973b,Bekenstein:1974b,
Bekenstein:1975b} to resolve certain thermodynamical paradoxes that
arise in the presence of the black holes and, in particular, to
preserve the universal applicability of the second law of
thermodynamics.  Soon after Bekenstein's proposal---based on their
classical, macroscopic behavior---the thermodynamic properties of 
black holes were formalized as the four laws of black hole
mechanics~\cite{Bardeen-Cart:1973}.  Specifically, it was argued that,
as the area theorem of classical general relativity closely resembles
the statement of the second law of thermodynamics, the area of the
black hole event horizon ($\cA_{_{\rm H}}$) can be interpreted as the
physical entropy associated with the black hole.  This association, in
turn, led to the identification of the surface gravity ($\kappa$) of 
the black hole (which, for a stationary black hole, is a constant all 
over the horizon) as the temperature of the hole.

The laws of black hole mechanics were placed on a firm footing when, a
year or two later, Hawking~\cite{Hawking:1974a,Hawking:1975a} showed
that, in the presence of quantum matter fields, a body that collapses
into a black hole emits thermal radiation at the temperature
\beq
T_{_{\rm H}} 
=\l(\frac{\hbar\, c}{k_{_{\rm B}}} \r)\, 
\l(\frac{\kappa}{2 \pi}\r)\,,
\eeq
where 
$\hbar$, $c$ and $k_{_{\rm B}}$ denote the Planck constant, the speed 
of light and the Boltzmann constant, respectively. The above Hawking 
temperature fixes the constant of proportionality between the temperature 
of the black hole and its surface gravity and, therefore, between the 
entropy and the area of the hole.  One finds that the entropy of black 
holes are given by the following Bekenstein-Hawking area law:
\beq
\label{eq:BHentropy}
\SBH 
= \l(\frac{k_{_{\rm B}}}{4}\r) \l(\frac{\cA_{_{\rm H}}}{\lPl^{2}}\r), 
\eeq
where $\lPl=\l(G\,\hbar/c^3\r)^{1/2}$ denotes the Planck length 
with $G$ being the Newton's constant.

Black hole entropy assumes considerable importance due to the fact
that it may provide us with an insight to the microscopic structure 
of the gravitational theory through the microcanonical, Boltzmann
relation $S = \l(k_{_{\rm B}} \ln \Omega\r)$, where $\Omega$ is the 
total number of quantum states that are accessible to a black hole 
that is described by a small set of classical parameters. The different
approaches that have been adopted in the literature to understand the
microscopic origin of black hole entropy can be broadly classified
into two categories. (i) Count the ``microstates'' by assuming a
fundamental structure like D-branes, spin networks or conformal
symmetry
\cite{Strominger-Vafa:1996a,Ashtekar-Baez:1997a,Carlip:2002a,Dasgupta:2005}.
(ii)
Associate the black hole entropy to the quantum fields propagating in
the fixed black hole spacetime. Count the microstates of these
quantum fields
\cite{'tHooft:1984a,Bombelli-Koul:1986,Srednicki:1993,Wald:1993a,
Frolov-Furs:1997a,Das-Shanki:2005a}.

Although none of the above approaches can be considered to be
complete; all of them --- within their domains of applicability --- by
counting certain microscopic states yield the semiclassical result
(\ref{eq:BHentropy}) in all spacetime dimensions $d \geq 3$.
However, all these approaches seem to lead to different sub-leading
corrections to the Bekenstein-Hawking entropy. For instance, (i) the
prefactor to the logarithmic corrections obtained using the
spin-networks and conformal symmetry
\cite{Kaul-Maju:2000a,Carlip:2000a,Govindarajan-Kaul:2001a,Gupta:2002pq,Hod:2004cd} 
are different from the one obtained using the statistical fluctuations
around thermal equilibrium \cite{Das-Maju:2001a}. (ii) the power-law
corrections obtained using the Noether charge approach
\cite{Wald:1993a} are different from those via entanglement of the
modes between inside and outside the horizon
\cite{Das-Shanki:2007b}. In other words, even though different degrees
of freedom lead to the universal Bekenstein-Hawking entropy --- {\it
quite naturally} --- they lead to different sub-leading terms.  This
indicates that the key to the understanding of the statistical
mechanical interpretation of Bekenstein-Hawking entropy may lie in the
origin of the sub-leading contributions.

Physically, it is natural to expect corrections to
(\ref{eq:BHentropy}): Bekenstein-Hawking entropy is a semi-classical
result and there are strong indications that this is valid for large
black holes [i.e. when horizon radius is much larger than Planck
length]. However, it is not clear, whether this relation will continue
to hold for the Planck size black holes. Besides, there is no reason
to expect that the Bekenstein-Hawking entropy to be the whole answer
for a correct theory of quantum gravity.

In this work, we calculate the higher order WKB contributions to the
Bekenstein-Hawking entropy from the brick-wall model
\cite{'tHooft:1984a,'tHooft:1996a,Gupta:2003pq}.  
We extend the zeroth-order ($\hbar^0$) 
WKB analysis to higher order and show that (i) The contribution to the 
entropy from the higher-order WKB modes is of the same order as the 
leading order WKB modes. In other words, our analysis shows that it 
may be incomplete to calculate the contribution only from the leading
order WKB modes. (ii) The brick-wall entropy $(\SBW)$ leads to generic
corrections to area of the form:
\beq
\SBW = \SBH + \GAH +  \FAH 
\log\l(\frac{\AHo}{\lPl^2}\r)\,,
\label{eq:mresult}
\eeq
where $\GAH,$ and $\FAH$ are polynomial functions of $\AHo$. In the
case of four-dimensions, we show explicitly that the brick-wall entropy 
(upto sixth-order) has the form given above with $\GAH = 0$. In the case 
of six-dimensions, $\GAH \neq 0$.  (iii) We show that, only in the case
of Schwarzschild, $\FAH$ is a constant.

The brick wall approach is a semi-classical approach, wherein the
background geometry is assumed to be a fixed classical background in
which quantum fields propagate.  The entropy of the black hole is
identified with the statistical mechanical entropy arising from a
thermal bath of quantum fields propagating outside the horizon.  The
entropy computed in this way turns out to be proportional to the area of
the horizon.  This approach has been very popular in obtaining the
leading order to the black hole entropy in different dimensions (for
an incomplete list of references, see Refs.
\cite{Paddy:1986a,Mann-Tara:1990a,Ghosh-Mitr:1994a,Demers-Lafr:1995,
Frolov-Furs:1995a,Kim-Kim:1996a,Kim-Kim:1996bp,Solodukhin:1996a,
Shen-Chen:1997a,Cognola-Lecc:1997a,Ho-Won:1997a,Solodukhin:1997a,
Frolov-Furs:1998a,
Ho-Kang:1998,Mukohyama-Isra:1998,Winstanley:2000a}).

The problem with the brick-wall model [as is the case of any
semiclassical approach] is that due to the infinite growth of density
of states close to the horizon, one has to impose ultra-violet cutoff
near the horizon and hence, the brick-wall entropy depends on the
cut-off scale. [See section (\ref{sec:BWall}), for discussion on various
aspects of brick-wall model.]  Clearly, this is an undesirable
feature. However, there are several advantages of the brick-wall model
over other approaches: (i) Unlike the Noether charge approach
\cite{Wald:1993a}, the brick-wall entropy depends only on the
kinematical properties of the metric close to the horizon and does not
depend on the dynamics. Hence, the brick-wall entropy (and the
corrections computed in this work) can directly be mapped to the horizon
properties.  In the case of Noether charge approach, since, such a
mapping is not possible, the power-law corrections does not provide
any new information about the statistical mechanical properties of
black hole entropy. (ii) Unlike entanglement entropy
\cite{Das-Shanki:2007b}, the brick-wall entropy can be computed
analytically for any spherically symmetric spacetimes to all
orders. Also, it is not possible to compute the entanglement entropy
for spacetime dimensions $d > 4$ --- the entropy is divergent.  (iii) In
the conformal field theory approach \cite{Carlip:2000a,Carlip:2002a},
the black hole horizon is treated as boundary. However, the vector
fields (which generate the symmetries) do not have a well-defined
limit at the horizon \cite{Dreyer-Ghos:2001a}. If one requires that
vector fields generating symmetries be smooth at the horizon, then the
central charge should be zero. In other words, the conformal field
theory analysis can only be performed close to the horizon like a
brick-wall.

The remainder of this paper is organized as follows.  
In the following section, we shall sketch some essential properties
of static, spherically symmetric black holes in arbitrary spacetime
dimensions.  
In Section~\ref{sec:BWall}, we shall discuss the assumptions and
approximations involved in evaluating the brick wall entropy, and
outline the procedure for extending the calculation to the higher 
orders (in terms of $\hbar$) in the WKB approximation. 
In Section~\ref{sec:BWCorr}, in addition to the zeroth order, we 
shall evaluate the contributions to the brick wall entropy of four 
dimensional black holes at the second, the fourth and the sixth 
orders (in terms of $\hbar$) in the WKB approximation.  
In Section~\ref{sec:HD-BW}, we extend the analysis to case of
black holes in six dimensions.  
In Section~\ref{sec:appl-BH}, we explicitly write down the results 
for a few specific black hole solutions in four and six dimensions.  
Finally, in Section~\ref{sec:discussion}, after a rapid summary 
of the results we have obtained, we shall discuss as to how the 
sub-leading contributions we have evaluated compare with the 
results obtained from the other approaches.

Before we proceed further, let us briefly outline the conventions 
and notations we shall adopt.  
We shall, in general, consider a $(D + 2)$-dimensional\footnote{A
comment on this uncommon notation may be in order.
We do {\it not}\/ work with two time coordinates!
We find it convenient to assume the spherically symmetric event 
horizon to be a $D$-dimensional sphere. 
The radial and the time coordinates make the dimension of 
spacetime to be~$(D+2)$.}, spherically symmetric, black hole 
spacetime. 
We shall work with the metric signature $(-,+,+,\cdots)$, and use 
the geometric units wherein $k_{_{\rm B}} = c = G = 1$.  
We shall denote the derivative of any function with respect to the 
radial coordinate~$r$ of the black hole by an overprime. 
The quantum field~$\Phi$ we shall consider will be a minimally 
coupled scalar field.

%%%%%%%%%%%%%%%%%%%%%%%%%%%%%%%%%%%%%%%%%%%%%%%%%%%%%%%%%%%%%%%%%%%%%%%%%%%%%%%

\section{Key properties of static, spherically symmetric black 
holes}\label{sec:ST}

Consider the following $(D + 2)$-dimensional static and 
spherically symmetric line element
\br
\label{eq:spher-tr}
ds^2
&=& -f(r)\, dt^2 + \frac{dr^2}{g(r)} 
+ r^2\, d\Omega_{_{D}}^2\,,\\ 
\label{eq:spher-tx}
&=&  f(r)\, \l[ -dt^2 + dx^2\r] + r^2\, d\Omega_{_{D}}^2\,,
\er
where $f(r)$ and $g(r)$ are arbitrary (but, continuous and
differentiable) functions of the radial coordinate $r$,
$d\Omega_{_{D}}^2$ is the metric on a $D$-dimensional unit 
sphere, and
\beq
x = \int \frac{dr }{\sqrt{f(r)\, g(r)}}
\label{eq:rela-xr}
\eeq
denotes the tortoise coordinate.  Throughout this work, we shall
assume that the line-element~(\ref{eq:spher-tr}) contains a
singularity (say, at $r = 0$) and {\it one},\/ non-degenerate, 
event horizon (located at, say, $r =\rHo$)\footnote{Since the 
event horizon is a null hypersurface, its location can be 
determined by the condition $\l(g^{\mu\nu}\, \pa_{\mu}N\, 
\pa_{\nu}N\r) = 0$, where $N$ is a scalar quantity that 
characterizes the hypersurface.  
For the line-element~(\ref{eq:spher-tr}), $N$ proves to be a 
function of the radial coordinate~$r$.
The above equation then leads to $g(\rHo) = 0$, and the roots of 
this {\it algebraic}\/ equation in turn determine the location of 
the horizon.}. 
But, we shall not assume any specific form of $f(r)$ or $g(r)$. 
In the rest of this section, we shall discuss some generic properties 
of the spacetime~(\ref{eq:spher-tr}) near the horizon at $r = \rHo$.

In almost all approaches that evaluate the entropy of spherically
symmetric black holes, their line-element close to the event horizon 
is approximated to be that of a Rindler spacetime (see, for instance, 
Ref. \cite{Frolov-Furs:1998a}).  
For the line-element~(\ref{eq:spher-tr}), the Rindler behavior 
near the horizon can be arrived at by first carrying out the 
following transformation of the radial coordinate:
\beq
\gamma = \l(\frac{1}{\kappa}\r) \sqrt{f},
\eeq
where $\kappa$ is a constant that denotes the surface gravity of the 
black hole and is defined as (see, for example, Ref.~\cite{Wald:1984-bk})
\beq
\kappa =  
\l[\sqrt{\frac{g(r)}{f(r)}}\, \l(\frac{f'(r)}{2}\r)\r]_{r=\rHo}\, .
\label{eq:kappa-def}
\eeq
In terms of the coordinate~$\gamma$, the 
line-element~(\ref{eq:spher-tr}) can be expressed as
\beq
ds^2 
= - \kappa^2\, \gamma^2\, dt^2 
+ 4 \, \l(\frac{f}{g}\r) \l(\frac{\kappa}{f'}\r)^2\,
d\gamma^2 + r^2\, d\Omega^2_D\,.
\eeq
Close to the horizon (i.e. near $r=r_{_{\rm H}}$), this  
line-element reduces to
\beq
ds^2 \to - \kappa^2\, \gamma^2\, dt^2 
+ d\gamma^2 + \rHo^2 \, d\Omega^2_D\,\label{eq:rind-met}
\eeq
which describes the Rindler spacetime with a horizon that is 
located at $\gamma = 0$. 
It should be stressed here that such a behavior is exhibited by 
all non-degenerate black hole horizons in all dimensions.

The above derivation of the Rindler line-element near the horizon 
is essentially equivalent to expanding the metric components~$f(r)$ 
and~$g(r)$ in~(\ref{eq:spher-tr}) about $\rHo$ up to the linear 
order in the Taylor series.
However, we find that, when evaluating the contributions to the 
brick wall entropy at the higher orders in the WKB approximation,
we need to expand the quantities $f(r)$ and $g(r)$ to higher 
orders as follows:
\begin{subequations}
\label{eq:fg-serh}
\br
f(r) &=& f'(\rHo)\, (r - \rHo) 
+ \l(\frac{f''(\rHo)}{2}\r)\, (r - \rHo)^2\\
& &\qquad\qquad\qquad\;\;
+\, \l(\frac{f'''(\rHo)}{6}\r)\, (r - \rHo)^3
+\ldots\,,\nn\\
g(r) 
&=& g'(\rHo)\, (r - \rHo) 
+ \l(\frac{g''(\rHo)}{2}\r)\, (r - \rHo)^2\\
& &\qquad\qquad\qquad\;\;
+\, \l(\frac{g'''(\rHo)}{6}\r)\, (r - \rHo)^3
+\ldots\,.\nn
\er
\end{subequations}
As we shall see, in four dimensions, in addition to the surface
gravity of the black hole, the corrections to the
Bekenstein-Hawking entropy~$\SBH$ also depend on the second
derivative of the metric evaluated at the horizon.
And, in six dimensions, we find that the sub-leading contributions 
to~$\SBH$ involve the third derivative of the metric as well.

Another quantity which we shall require in our calculations is the 
proper or the coordinate invariant distance of the brick wall from 
the horizon. 
The proper radial distance to the brick wall, say, $h_{c}$, that is 
located at $r=h$ is given by
\beq
h_c = \int\limits_{\rHo}^{\rHo + h} \frac{dr}{\sqrt{g(r)}}\,.
\label{eq:defhc}
\eeq
On using the expansion~(\ref{eq:fg-serh}) for~$g(r)$ up to the
second order in this integral, we obtain the following relation 
between $h$ and $h_{c}$:
\beq
h^{1/2} = \sqrt{\frac{2\, g'(\rHo)}{g''(\rHo)}}\; 
\sinh \l[\sqrt{\frac{g''(\rHo)}{2}}\l(\f{h_c}{2}\r)\r]\,.
\eeq
For small $h_{c}$, this relation simplifies to
\beq
h_c = \sqrt{\frac{4 h}{g'(\rHo)}} \, ,
\label{eq:defhc-rind}
\eeq
and, for convenience, we shall use this expression for the proper
distance to the brick wall.

%%%%%%%%%%%%%%%%%%%%%%%%%%%%%%%%%%%%%%%%%%%%%%%%%%%%%%%%%%%%%%%%%%%%%%%%%%%%%%%

\section{Extension of the brick wall model to higher 
orders in the WKB approximation}\label{sec:BWall}

In this section, after a rapid sketch of the assumptions and
approximations that are involved in evaluating the black hole 
entropy using the brick wall model, we go on to outline the 
procedure for computing the brick wall entropy at the higher 
orders in the WKB approximation.

%%%%%%%%%%%%%%%%%%%%%%%%%%%%%%%%%%%%%%%%%%%%%%%%%%%%%%%%%%%%%%%%%%%%%%%%%%%%%%%

\subsection{Basic assumptions}\label{sec:BWapprox}

There are two crucial assumptions in the brick wall approach to black
hole entropy.  The first assumption concerns the modeling of the
microscopic origin of the black hole entropy, and the second is
regarding the handling of the divergences that arise close to the
event horizon.

As we have mentioned before, the brick wall model is a semi-classical
approach wherein the black hole is assumed to be described by a fixed
classical geometry.  It is further assumed that the black hole is in
equilibrium with a thermal bath of quantum matter fields at the
Hawking temperature. Moreover, it is the canonical entropy (actually,
a specific component) of the quantum matter fields that are
propagating outside the black hole horizon that is identified to be
the entropy of the black hole.

In the process of calculating the canonical entropy of a matter field
outside the black hole horizon, we need to evaluate the density of
states of the field.  However, one finds that, due to the infinite
blue shifting of the modes in the vicinity of the event horizon, the
density of states actually diverges.  This divergence is regulated in
the model by introducing a cut-off by hand above the horizon.  The
cut-off---popularly referred to as the brick-wall---is basically a
static, spherical mirror at which the matter fields are assumed to
satisfy, say, the Dirichlet boundary conditions.  One finds that the
leading component of the brick wall entropy diverges as $h_{c}^{-2}$, 
where $h_{c}$ is the proper distance to the brick wall defined 
in Eq.~(\ref{eq:defhc}).  (The other component
is essentially a volume dependent term that arises even in flat
space.)  It is this contribution that is identified to be the entropy
of the black hole.  Moreover, a specific choice for the cut-off $h_c$
has to be made (this depends on the number of fields, the dimension of
the spacetime, etc., but is generally of the order of the Planck
length ${\ell_{_{\rm Pl}}}$), in order to reproduce the
Bekenstein-Hawking area law~(\ref{eq:BHentropy}).  As we mentioned,
the area law~(\ref{eq:BHentropy}) has been recovered in this approach
for a variety of black hole spacetimes and matter
fields~\cite{Paddy:1986a,Mann-Tara:1990a,Ghosh-Mitr:1994a,Demers-Lafr:1995,
Frolov-Furs:1995a,Kim-Kim:1996a,Kim-Kim:1996bp,Solodukhin:1996a,
Shen-Chen:1997a,Cognola-Lecc:1997a,Ho-Won:1997a,Solodukhin:1997a,
Frolov-Furs:1998a,
Ho-Kang:1998,Mukohyama-Isra:1998,Winstanley:2000a,Jassal:2007}.

%%%%%%%%%%%%%%%%%%%%%%%%%%%%%%%%%%%%%%%%%%%%%%%%%%%%%%%%%%%%%%%%%%%%%%%%%%%%%%%

\subsection{Essential approximations}

Two approximations turn out to be essential to make the computation of
the brick wall entropy tractable.  The first approximation is required
in evaluating the density of states of matter fields around black
holes, and the second involves expanding the metric near the event
horizon.
 
As we pointed out above, in order to evaluate the brick wall entropy,
one needs to evaluate the density of states of matter fields around
black holes.  However, apart from some lower dimensional cases, the 
density of states cannot be evaluated exactly.  As a result, in the 
brick wall model, the density of states is usually evaluated at the 
leading order in $\hbar$ in the WKB approximation.

Moreover, barring a few special cases, one finds that, even after the
WKB approximation, the brick wall entropy cannot be evaluated exactly.
Recall that the dominant contribution to the entropy arises due to the
modes close to the horizon.  Motivated by this feature, one Taylor
expands the metric functions $f(r)$ and $g(r)$ near the horizon in
order to obtain a closed form expression for the brick wall entropy.

%%%%%%%%%%%%%%%%%%%%%%%%%%%%%%%%%%%%%%%%%%%%%%%%%%%%%%%%%%%%%%%%%%%%%%%%%%%%%%%

\subsection{The methodology}\label{sec:BWmeth}

Having discussed the assumptions and approximations involved in the 
brick wall approach, in the remainder of this sub-section, we shall 
outline the procedure for evaluating the brick wall entropy at the
higher orders in the WKB approximation.

The key assumption of the brick wall model, as we have pointed 
out above, is that the black hole is in equilibrium with a bath 
of thermal radiation at the Hawking temperature of the hole.
The free energy $F$ of a scalar field at the inverse 
temperature~$\beta$ is given by  (see, for example, 
Ref.~\cite{'tHooft:1984a})
\br
F &=& \l(\frac{1}{\beta}\r)\, 
\int\limits_0^{\infty} dE\, \l(\f{d\Gamma(E)}{dE}\r)\, 
\ln\l[1 - \exp-(\beta E)\r]\,,\nn\\
&=& - \int\limits_0^{\infty} dE\, 
\l(\frac{\Gamma(E)}{\exp(\beta E) - 1}\r)\,,\label{eq:def-F}
\er
where $\Gamma(E)$ denotes the total number of modes of the field
with energy less than $E$. We have integrated the first of 
the above equation by parts to arrive at the second and have 
assumed that the boundary term vanishes. The canonical entropy 
associated with the free energy $F$ is given by
\beq
S_{_{\rm C}}(\b)
=  \beta^2 \, \l(\frac{\pa F}{\pa \beta}\r)\,,
\label{eq:CE-entropy}
\eeq
and, it is this entropy, evaluated at the Hawking temperature, that
will be identified to be the entropy of the black hole. 

\begin{widetext}
Consider a massive and minimally coupled scalar field~$\Phi$
that is propagating in the the line-element~(\ref{eq:spher-tr}). 
Such a field satisfies the differential equation
\beq
\label{eq:eomPhi}
\l(\Box-m^2\r)\,\Phi=0,
\eeq
where $m$ denotes the mass of the field.
The rotational symmetry of the line-element~(\ref{eq:spher-tr}) 
allows us to decompose the normal modes $u_{_{E\ell m_{i}}}$ of 
the field $\Phi$ as follows (see, for instance, 
Ref.~\cite{Frolov-Mazz:1989}):
\beq
u_{_{E\ell m_{i}}}(x^{\mu}) 
= \l(\frac{R(r)}{r^{D/2}\,
G^{1/2}(r)}\r)\, Y_{\ell m_i}(\theta,\phi_i)\;\, 
e^{-\l(iEt/\hbar\r)}\,,
\label{spher-exp}
\eeq
where $E$, $\ell$ and $m_{i}$ (with $i \in [1, (D - 1)]$) are the 
energy, angular momentum and the azimuthal angular momenta 
associated with the modes, respectively, the quantity $G(r)$ is 
given by
\beq
\label{eq:defG}
G(r) = \sqrt{f(r)\; g(r)}\,,
\eeq
and $Y_{lm_i}(\theta,\phi_i)$ denote the hyper-spherical harmonics.
On substituting the mode~(\ref{spher-exp}) in the equation 
of motion~(\ref{eq:eomPhi}) and using the properties of the 
hyper-spherical harmonics, we find that the function $R(r)$
satisfies the differential equation
\beq
\label{eq:eff-Sch}
R''(r) + \l[\frac{V^2(r)}{\hbar^2} - \Delta(r) \r]\, R(r) = 0\, ,
\eeq
where the quantities $V^2(r)$ and $\Delta(r)$ are given by
\begin{subequations}
\br
V^{2}(r)
&=& \l(\frac{1}{G^2(r)}\r)\, \l(E^2 - f(r)\, 
\l[m^2 + \l(\frac{\ell\, (\ell + D - 1)\, 
\hbar^2}{r^2}\r)\r]\r)\,,\label{eq:defV}\\
\Delta(r) 
&=& \l(\frac{G''(r)}{2\, G(r)}\r) - 
\l(\frac{G'^{2}(r)}{4\, G^{2}(r)}\r)
+  \l(\frac{D}{2 r}\r)\, \l(\frac{G'(r)}{G(r)}\r)
+ \l(\frac{D\, (D - 2)}{4\, r^2}\r)\,.\label{eq:defDel}
\er
\end{subequations}

The total number of modes~$\Gamma(E)$ of the field $\Phi$ with 
energy less than $E$ can be evaluated exactly if the solution 
to the differential equation~(\ref{eq:eff-Sch}) can be written 
down explicitly.
However, apart from some simple $(1+1)$-dimensional 
example\cite{Shen-Chen:1997a}, it proves to be difficult to obtain an exact 
analytical solution for the function~$R(r)$.
As a result, the WKB approximation is almost always resorted to in
the literature
\cite{Paddy:1986a,Mann-Tara:1990a,Ghosh-Mitr:1994a,Demers-Lafr:1995,
Frolov-Furs:1995a,Kim-Kim:1996a,Kim-Kim:1996bp,Solodukhin:1996a,
Shen-Chen:1997a,Cognola-Lecc:1997a,Ho-Won:1997a,Solodukhin:1997a,
Frolov-Furs:1998a, Ho-Kang:1998,Mukohyama-Isra:1998,Winstanley:2000a}, 
and it is the leading order WKB solution for $R(r)$ that is utilized 
to evaluate the number of 
states~$\Gamma(E)$, and the resulting free energy~$F$ and the 
entropy of~$S_{_{\rm C}}$ of the quantum field.
Our goal here is to extend the analysis to the higher orders in 
the WKB approximation.

Let us begin by expressing the function~$R(r)$ in the following WKB 
form:
\beq
R(r) = \l(\frac{c_0}{\sqrt{P(r)}}\r)\,
\exp\l[\frac{i}{\hbar} \int\limits^{r} d{\tilde r}\, 
P({\tilde r})\r]\,,\label{eq:WKBans}
\eeq
where $c_0$ is a constant.
On substituting this expression in Eq.~(\ref{eq:eff-Sch}), we find 
that the function~$P(r)$ satisfies the differential equation
\beq
\l(\frac{1}{\hbar^2}\r)\, \l[P^2(r) - V^2(r)\r]
= \l(\frac{3}{4}\r) \l(\frac{P'(r)}{P(r)}\r)^2 - 
\l(\frac{1}{2}\r)\, \l(\frac{P''(r)}{P(r)}\r) - \Delta(r)\,.
\label{eq:HJeq}
\eeq
Let us now expand the function~$P(r)$ in a power series in~$\hbar^2$ 
as follows (see, for instance, Ref.~\cite{Bender-Orza:1978}):
\beq
P(r) = \sum_{n=0}^{\infty} \hbar^{2 n}\; P_{2 n} (r)\,.
\label{eq:for-Pexp}
\eeq
On substituting this series in the differential 
equation~(\ref{eq:WKBans}) and collecting the terms of a given 
order in~$\hbar^2$, we obtain following expressions for $P_{2n}(r)$ 
upto~$n=3$:
\begin{subequations}
\label{eq:formal-Ps}
\br
P_0(r) 
&=& \pm V(r) 
= \pm\, \l(\frac{1}{G(r)}\r)\, \l[E^2 - f(r)\,
\l(m^2 + \l[\frac{\ell\, (\ell + D - 1)\, 
\hbar^2}{r^2}\r]\r)\r]^{1/2}\,,\label{eq:for-P0}\\
P_2(r) 
&=&  \l(\frac{3}{8 P_0(r)}\r)\, \l(\frac{P_0'(r)}{P_0(r)}\r)^2 
- \l(\frac{4\, P_0''(r)}{{P_0}^2(r)}\r)
- \l(\frac{\Delta (r)}{2\, P_0(r)}\r)\,,\label{eq:for-P2}\\
P_4(r) 
&=& -\l(\frac{5\, {P_{2}}^{2}(r)}{2\, V(r)}\r) 
-\l(\frac{4\, {P_2}(r)\, \Delta (r) + P_{2}''(r)}{4\, V^{2}(r)}\r)
+\l(\frac{3\, {P_2}'(r)\, V'(r)-{P_2}(r)\, 
V''(r)}{4\, V^{3}(r)}\r)\,,\label{eq:for-P4}\\
P_{6}(r) 
&=& -\l(\frac{5\, P_{2}(r)\, P_{4}(r)}{V(r)}\r)
-\l(\frac{8\, {P_{2}}^3(r)+4\, P_{4}(r)\, \Delta(r) 
+P_4''(r)}{4\, V^2(r)}\r) 
- \l(\frac{\Delta(r) {P_2}^2(r)}{2\, V^3(r)}\r)\nn\\ 
\label{eq:for-P6}
& &\qquad\qquad\qquad\qquad\;\,
-\l(\frac{2\, P_{2}''(r)\, P_{2}(r)+2\, P_{4}(r)\, V''(r)
- 3\, {P_{2}'}^{2}(r) - 6\, P_{4}'(r)\, V'(r)}{8\, V^3(r)}\r)\,.
\er
\end{subequations}
\end{widetext}

Note that the function $P_0(r)$ is related {\it algebraically}\/ 
to the quantities $V(r)$ and $\Delta(r)$.
It is evident that the higher order functions 
$P_{2 n}(r)$  (with $n > 0$) can be expressed in terms of the 
functions at the lower orders and their derivatives and, eventually, 
in terms of the function~$P_{0}(r)$.

On using the series expansion~(\ref{eq:for-Pexp}) in the standard
semiclassical quantization procedure~\cite{'tHooft:1984a} , we
can express the total number of states~$\Gamma(E)$ of the field
with energy less than $E$ as follows:
\beq
\Gamma(E)=\sum_{n=0}^{\infty}\, \Gamma_{2n}(E),
\label{eq:Gamma}
\eeq
where we have defined $\Gamma_{2 n}(E)$ as
\br
\label{eq:Gamma-order}
\Gamma_{2 n}(E)\!\!
&=&\!\! \l(\frac{\hbar^{2 n -1}}{\pi}\r)\!\! 
\int\limits_{_{\rHo + h_c}}^{L}\! dr 
\int\limits_{0}^{\ell_{max}}\! d\ell\; 
\l(2 \ell + D - 1\r)\nn\\
& & \qquad\qquad\qquad\qquad\qquad
\times\; 
{\cal W}(\ell)\; P_{2 n}(r)\,,\qquad\;\;
\er
with the quantity ${\cal W}(\ell)$ being given by
\beq
\label{eq:Weightl} 
{\cal W}(\ell) 
= \l(\frac{(\ell + D - 2)!}{(D - 1)! \, \, \ell!}\r)\,. \\
\eeq
It should be mentioned that, in the above expression for 
$\Gamma_{2 n}(E)$, we have approximated the sum over the angular 
quantum numbers~$\ell$ as an integral with a degeneracy factor 
${\cal W}(\ell)$.
Such an approximation is often made in the literature, and the
approximation is considered to be valid since the separation 
between the states are expected to be small~\cite{Kim-Kim:1996bp}.
Moreover, the upper limit $\ell_{_{max}}$ on the integral 
over $\ell$ is a function of energy $E$ of the mode and the
radial coordinate $r$, and it has to be chosen such that 
$P_{0}(r)$ is real\footnote{Actually, the limits have to be 
chosen such that $P_{2n}(r)$ are real for all $n$. 
However, since, for $n>0$, the functions $P_{2n}(r)$ can be 
expressed in terms of~$P_{0}(r)$ and  the real functions~$V(r)$ 
and $\Delta(r)$, when $P_{0}(r)$ is real, $P_{2n}(r)$ are real 
as well.
Therefore, the limits on $\ell$ proves to be the same for all~$n$.}.
Furthermore, the lower limit on the integral over radial coordinate, 
viz. $h_c$, is the invariant thickness of the `brick-wall' defined 
in~(\ref{eq:defhc}), and the upper limit ${L}$ is the infra-red 
cutoff which we shall assume to be much larger than the horizon 
radius.

A few clarifying remarks are in order at this stage of our 
discussion.
In the semi-classical quantization of, say, a one-dimensional
non-relativistic quantum particle, the integral over the 
coordinate will be carried out over the range wherein $P_{0}$ 
is real~\cite{Bender-Orza:1978}).   
In the case of bounded systems, these limits will prove to 
be the turning points of the potential, whereas in the case 
of potential barriers the limits will be between one of the
turning points and infinity.  
In the context of black holes, the effective potential turns out 
to be a barrier and the integral over the radial coordinate is to 
carried out between the event horizon of the black hole (which is 
an infinity in terms of the tortoise coordinates) and the first 
turning point that is located on the barrier.
%\footnote{For instance, 
%in the case of four-dimensional Schwarzschild black hole, the 
%Regge-Wheeler potential~(\ref{eq:rwpot}) has two turning points 
%which are located far from the event horizon as well as radial 
%infinity~\cite{Iyer-Will:1986}.}. 
But, one finds that, most of the contribution to the density of 
states of the quantum field arises due to the modes close to
the event horizon of the black hole, while the upper limit located
on the barrier leads to a volume dependent contribution to the 
entropy. 
As a result, the contribution to the number of states and the free
energy and the entropy of the quantum field due to the upper 
(infra-red) limit is usually ignored in the literature. 

We should emphasize the point that, apart from replacing the sum 
over $\ell$ by an integral, we have not made any approximations 
until now.
Hereafter, we shall make two approximations that we had discussed 
is some detail in the last subsection.  
Firstly, we shall approximate the line-element~(\ref{eq:spher-tr}) 
near the event horizon of a spherically symmetric black hole to be 
that of Rindler spacetime, viz. Eq.~(\ref{eq:rind-met}).
It should be pointed out that such an approximation is always made
in the literature to arrive at closed form expressions for the free
energy and the entropy of the quantum field.
Secondly, we shall truncate the series~(\ref{eq:for-Pexp}) at a 
particular order (we shall work until the sixth order in $\hbar$),
and evaluate the density of states and the associated free energy 
and the entropy of the quantum field around the black hole. 
It is important to note that, in the literature, it is only the 
leading term in the series~(\ref{eq:Gamma}) that has {\it always}\/
been taken into account ignoring the higher orders when evaluating 
the brick-wall entropy\footnote{The higher order WKB procedure 
we use is different compared to the approach used in the 
quasi-normal modes \cite{Iyer-Will:1986,Iyer:1986b,Konoplya:2003a}. 
In Ref. \cite{Iyer-Will:1986}, the Regge-Wheeler potential is 
expanded around the maxima, and the modes close to the 
maxima are matched to the one close to the horizon.}.

In the following two sections, we shall evaluate the contributions
to the brick wall entropy at the higher order for four and six 
dimensional black holes, respectively.
As we shall see, the contributions to the entropy from the higher 
orders turn out to be of the same order as the leading order in 
the WKB approximation.  
In other words, it may be incomplete to calculate the contribution 
only from the leading term in the WKB expansion. 
Moreover, we show that the brick wall entropy leads to generic 
corrections to Bekenstein-Hawking entropy~(\ref{eq:BHentropy}). 
For instance, in the case of four-dimensions, we find that the 
brick-wall entropy has the form:
\beq
S_{_{\rm BW}}^{\rm (4D)} 
= S_{_{\rm BH}} + \FAHFD\, 
\log\l(\frac{{\cal A}_{H}}{\ell_{_{\rm Pl}}^2}\r),
\eeq
where $\FAHFD$ depends on the surface gravity and the second 
derivative of the metric at the horizon.
%$\FAHFD$ is a constant only for Schwarzschild black hole. 

Before we proceed with the calculations, there is yet another point 
concerning the WKB approximation at the sub-leading orders that we 
need to discuss.
As we mentioned above, the limits on the integral over $\ell$ has 
been chosen such that $P_{0}(r)$ is real.
This condition essentially identifies the turning points of the
potential.
Notice that, in Eq. (\ref{eq:formal-Ps}), all the higher order 
WKB terms---i.e. $P_{2n}(r)$ for $n>0$---contain $P_{0}(r)$ in 
the denominator.
Obviously, these functions will diverge at the turning points,
or equivalently, at the upper limits $\ell$. 
Such a divergence is a well-known feature of the WKB 
approximation at the higher orders\cite{Bender-Orza:1978}, and 
we shall devise a systematic procedure to isolate these 
divergences.
We shall outline this procedure in the next section and relegate
some of the details to Appendix~(\ref{app:Leibnitz}).

\section{Higher order contributions  
in four dimensions}\label{sec:BWCorr}

In this section, we shall evaluate the brick wall entropy for 
spherically symmetric, four dimensional black holes by 
considering the contributions up to the $n=3$ term in the 
series expansion~(\ref{eq:Gamma}) for the number of states of 
the quantum field.
For simplicity, we shall consider here the case of $f(r) =g(r)$ 
in the line-element~(\ref{eq:spher-tr}) and restrict ourselves to 
a massless scalar field (i.e. $m=0$).
In Appendix~(\ref{app:corr-genmetric}), we shall extend the second
order results we obtain in this section for the general case wherein 
$f(r)\neq g(r)$ and, in Appendix~(\ref{app:corr-mass}), we extend the
analysis to a massive field, but restrict ourselves to the case $f(r) 
=g(r)$.

%%%%%%%%%%%%%%%%%%%%%%%%%%%%%%%%%%%%%%%%%%%%%%%%%%%%%%%%%%%%%%%%%%%%%%%%%%%%%%%

\subsection{Second order}\label{sec:4D-IIorder}

Let us now evaluate the contribution due to the $n=1$ term in the series~(\ref{eq:Gamma}).
For $f(r)=g(r)$, we find that the expression~(\ref{eq:for-P2})
for second order `momentum' $P_2(r)$ can be written as 
\br
P_2(r) 
&=& \l(\frac{P^{(0)}_2 (r)}{{\cal G}({\cal E},r)}\r)
+ \lambda(r)\,
\l(\frac{P^{(1)}_{2}(r)}{{\cal G}^{3}({\cal E},r)}\r)\nn\\
& &\qquad\qquad\quad
+\, \lambda^{2}(r)\, 
\l(\frac{P^{(2)}_{2}(r)}{{\cal G}^{5}({\cal E},r)}\r)\,,
\label{eq:p2-f=g}
\er
where the functions $P^{(0)}_2 (r)$, $P^{(1)}_2 (r)$ and
$P^{(2)}_2(r)$ are given by
\br
\!\!
P^{(0)}_2 (r) 
\!\!&=& -\l(\frac{g'}{2\, r}\r)\,,\nn\\
\!\!P^{(1)}_2 (r) 
\!\!&=&\!\! \l(\!\frac{{g'}^2(r)}{8\, g(r)^2}\!\r)
-\l(\!\frac{3\, g'(r)}{4\, r\, g(r)}\!\r)
+\l(\!\frac{g''(r)}{8\, g(r)}\!\r)+\l(\!\frac{3}{4\, r^2}\!\r)\,,\nn\\
\label{eq:P2-0-2-f=g}
\!\!
P^{(2)}_{2}(r) 
\!\!&=&\!\! \l(\frac{5}{32}\r) \l(\frac{g'(r)}{g^2(r)}\r)^{2}\!
-\!\l(\frac{5\, g'(r)}{8\, r\, g(r)}\r)+\l(\frac{5}{8\, r^2}\r)\,,\;\;
\er
and, for convenience, we have defined
\beq
\CGE =\l[{\cal E}-\lambda(r)\r]^{1/2}\label{eq:def-CGE}
\eeq
with  ${\cal E}=E^2$ and $\lambda(r)$ being given by
\beq
\label{eq:def-lambda}
\lambda(r) = \l[\ell\, (\ell + 1)\, \hbar^2\r] \, 
\l(\frac{g(r)}{r^2}\r)\,.
\eeq

We now need to substitute the above expression for $P_{2}(r)$ 
in Eq.~(\ref{eq:Gamma-order}) and evaluate the number of modes 
$\Gamma_{2}$ with the upper limit $\ell_{\rm max}$ on the integral 
over $\ell$ being determined by the condition that the term
${\cal G}({\cal E},r)$ vanishes.
Clearly, the integral over $\ell$ will diverge in such a case. 
In order to isolate the finite contribution due to these higher 
order WKB modes, it is necessary that we follow a systematic 
procedure.  
The procedure we shall adopt is as follows.  
We shall first rewrite all the terms containing inverse powers of 
${\cal G}({\cal E},r)$ in terms of derivatives of ${\cal E}$ as 
follows:
\begin{subequations}
 \label{eq:deriv-E}
\begin{eqnarray}
\label{eq:deriv-Ea}
\l(\frac{1}{{\cal G}({\cal E},r)}\r)
&=& 2\; \l(\frac{\partial {\cal G}({\cal E},r)}{\partial {\cal E}}\r), \\
%%%%
\label{eq:deriv-Eb}
\l(\frac{1}{{\cal G}^{3}({\cal E},r)}\r)
&=& -4\; \l(\frac{\partial^{2} 
{\cal G}({\cal E},r)}{\partial {\cal E}^{2}}\r), \\
%%%
\label{eq:deriv-Ec}
\l(\frac{1}{{\cal G}^{5}({\cal E},r)}\r)
&=& \l(\frac{8}{3}\r)\,
\l(\frac{\partial^{3} {\cal G}({\cal E},r)}{\partial {\cal E}^{3}}\r).
\end{eqnarray}
\end{subequations}
Then, before evaluating the $\ell$ integral, we shall make use of
the Leibnitz's rule, viz.
\begin{eqnarray}
& &\!\!\!\!\!\!\!\!\!\!\!\!\!\!\!\!\!\!\!\!\!\!
\frac{\partial}{\partial x} \!\! 
\int\limits_{a(x)}^{b(x)}\! dt\, f[x,t]\nn\\
& &\!\!\!\!\!\!\!\!\!\!\!\!
=\, f[x,a(x)]\, \l(\frac{da(x)}{dx}\r) - f[x,b(x)]\, 
\l(\frac{db(x)}{dx}\r)\nn\\
& &\qquad\qquad\qquad\qquad+ \, \, 
\int\limits_{a(x)}^{b(x)}dt\, \l[\frac{\partial f(x,t)}{\partial x}\r] .
\label{lebtz}
\end{eqnarray}
and interchange the order of differentiation and integration over the
energy $E$ and $\ell$.  
When we do so, we find that the divergences occur at the turning 
point. We have checked the procedure up to
the $6^{\rm th}$-order WKB modes and, indeed, systematically separates
the non-divergent part from the divergent.  For completeness, in
Appendix (\ref{app:Leibnitz}), we give the details of the above
procedure.
(The procedure involves
calculating the contour integral around the branch cut that joins the
turning points.  For details, see Sec. (10.7) in
Ref. \cite{Bender-Orza:1978}.)

Having obtained the non-divergent part of the mode-functions as a
function of $E$, our next step is to evaluate the contribution of
these modes to the density of states $\Gamma_2{(E)}$. Using the
general expression (\ref{eq:Gamma-order}), we have
\begin{eqnarray}
\Gamma_2{(E)}=\frac{\hbar}{\pi} \int\limits_{\rHo +h}^{L} dr
\int\limits_{0}^{\ell_{max}}d\ell \, (2\ell+1) \, P_{2}(r) \, .
\end{eqnarray}
Substituting for $P_2(r)$ from Eq. (\ref{eq:p2-f=g}) and using the
relations (\ref{eq:deriv-E}), we get
%
%\begin{widetext}
\begin{eqnarray}
 \hbar\Gamma_2{(E)}&=&
\frac{1}{\pi} \int\limits_{_{\rHo +h}}^{L} \!\! dr \frac{r^2 P^{(2)}_0(r)}{2}
\int\limits_{0}^{\cE} d\lambda \, \frac{\partial \CGE}{\partial \cE}  \\
%%%
&-&\frac{1}{\pi} \int\limits_{_{\rHo +h}}^{L} dr \, r^2 P^{(2)}_1 (r)
\int\limits_{0}^{\cE}d\lambda \, \, \lambda \, 
\frac{\partial^2 \CGE}{\partial \cE^2} \nn \\
%%%%%%%
&+&\frac{1}{\pi} \int\limits_{_{\rHo +h}}^{L} \!\! dr \, \frac{3 r^2
P^{(2)}_2 (r)}{2}
\int\limits_{0}^{\cE}d\lambda \, \, \lambda^2 \, 
\frac{\partial^3 \CGE}{\partial \cE^3} \, . \nn
\label{gamma2}
\end{eqnarray}
%\end{widetext}

Using the Leibniz rule (\ref{lebtz}) and following the steps discussed
in Appendix (\ref{app:Leibnitz}), we get
{\small 
\beq 
\Gamma_2 (E) = \frac{E}{\hbar \pi} \int\limits_{_{_{\rHo
+h}}}^{L} \!\!\!\!\!  dr \l[ \frac{1}{3}-\frac{4 r g'(r)}{3 g(r)}+r^2
\left\{\frac{g'(r)^2}{3 g(r)^2}-\frac{g''(r)}{2 g(r)}\right\} \r] \, .
\eeq } 
Following points are worth noting regarding the above expression: (i)
In the case of leading order WKB modes, the density of states goes as
$E^3$ [see Eq. (\ref{eq:GammaSF})]. However, for the second-order WKB
modes the density of states scales as $E$. (ii) As in the
leading-order, most of the contributions to the entropy come close to
the horizon. (iii) The expression for the density of state
(\ref{eq:Gamma-fneqg}) for the general spherically symmetric
spacetime is same as for the special case discussed in this
section. Hence, the dependence on the entropy with area is identical
to the special case discussed in this section.

Substituting the above expression in \eq{eq:def-F}, and integrating
over $E$, the free-energy is
{\small
\beq
F_2 = -\frac{\pi}{6 \hbar \beta^2} \int\limits_{_{\rHo +h}}^{L} \!\!\!\!\! 
dr \left[\frac{1}{3}-\frac{4 r g'(r)}{3 g(r)}+r^2 \left(\frac{g'(r)^2}{3
g(r)^2}-\frac{g''(r)}{2 g(r)}\right)\right].
\eeq
}
Using the relation (\ref{eq:CE-entropy}), the entropy is given by
{\small
\beq
S_2 = \frac{\pi}{3 \hbar \beta} \int\limits_{\rHo +h}^{L}dr
\left[\frac{1}{3}-\frac{4 r
g'(r)}{3 g(r)}+r^2 \left(\frac{g'(r)^2}{3 g(r)^2}-\frac{g''(r)}{2
g(r)}\right)\right]. \label{en2}
\eeq
}
As mentioned above, maximum contribution to the entropy is from the
modes close to the horizon. Hence, using the expansion
(\ref{eq:fg-serh}) close to the horizon and the definition of
surface gravity (\ref{eq:kappa-def}), we get,
\begin{eqnarray}
S_2 = \frac{1}{9}\frac{\rHo^2}{h^{2}_{c}} 
- \l[\frac{g''(\rHo) \rHo^2}{72} + \frac{\kappa}{9} \rHo\r]
\log \l(\frac{\rHo^2}{h_c^2}\r) 
\label{en2I}
\end{eqnarray}
where $h_c$ is given by Eq. (\ref{eq:defhc-rind}).  
This is the first result of this paper,
regarding which we would like to stress the following points:
\begin{enumerate}
\item The dependence of the entropy on area (from the second-order 
WKB modes) is similar to that from the zeroth order WKB modes 
(\ref{eq:EntrF}). Also the contribution to the entropy 
from the second order WKB modes contribute more as 
compared to the leading order WKB modes. This result has 
two immediate consequences:

(a) To associate the brick-wall entropy to $\SBH$ it is {\it
necessary} to calculate all the higher order WKB mode contribution to
the brick-wall entropy.

(b) The sub-leading corrections (at the zeroth and second order WKB)
depend only on the surface gravity and second derivative of the metric
functions. They are of the form $ \FAH \log(\AHo/h_{c}^2)$.  To confirm
the generic structure for higher-order, in the next two subsections we
evaluate fourth and sixth order contributions to the brick-wall
entropy\footnote{It should be noted that, in the case of sixth order
WKB modes, the integral over $E$ is divergent near $E \to 0$. However, 
the near-horizon contribution of the entropy in identical
to the one obtained in this subsection. The fourth order WKB modes do 
not contribute to the brick-wall entropy.}.
\item If the surface gravity is inversely proportional to horizon
radius and $g''(\rHo)$ is inversely proportional to the square of the
horizon radius, then second term in the RHS of (\ref{en2I}) is a
constant. In this case, the corrections to $\SBH$ are purely
logarithmic and does not contain any power-law dependence. This 
uniquely corresponds to Schwarzschild spacetime.

In the case of Schwarzschild, we have
\beq
f(r) = g(r) = 1 - \frac{2 M}{r}
\label{eq:4DSch}
\eeq
where $M$ is the mass of the black hole. The horizon is at $\rHo = 2
M$, $\kappa = 1/(4 M)$ and $g''(\rHo) = -1/(2 M^2)$. Substituting the
above expressions in Eq. (\ref{en2I}), we get
\beq
S_2 =\frac{4}{9} \frac{M^2}{h^{2}_c} 
-\frac{1}{36}\log\l(\frac{\rHo^2}{h_c^2}\r) .
\eeq
This result shows that, at least, in the zeroth and second order,
there are no power-law corrections to $\SBH$ for the four-dimensional
Schwarzschild black hole, while, for all other black holes --- since
$\kappa$ and $g''(r)$ has a more non-trivial structure -- there are
power-law corrections to the Bekenstein-Hawking entropy. This leads to
the following conclusion: {\it The power-law corrections to the entropy
occur for any non-vacuum solutions.} In Sec. (\ref{sec:appl-BH}) we
obtain the entropy for some known black hole solutions.
\end{enumerate}

%%%%%%%%%%%%%%%%%%%%%%%%%%%%%%%%%%%%%%%%%%%%%%%%%%%%%%%%%%%%%%%%%%%%%%%%%%%%%%%

\subsection{Fourth order}\label{sec:4D-IVorder}
Using the expression (\ref{eq:for-P4}), we get,
\br
P_4(r) &=& \frac{P^{(0)}_4 (r)}{\CGEV{3}} 
+ \frac{\lambda(r) P^{(1)}_{4}(r)}{\CGEV{5}}  
+ \frac{\lambda^{2}(r) P^{(2)}_{4}(r)}{\CGEV{7}} \nn \\
\label{eq:p4-f=g}
& & + \, \, \frac{\lambda^{3}(r) P^{(3)}_{4}(r)}{\CGEV{9}} 
+ \frac{\lambda^{4}(r) P^{(4)}_{4}(r)}{\CGEV{11}} \, ,
\er
where, the complete form of $P^{(i)}_{4}(r)$ [where $i = 0 \cdots 4$]
are given in Appendix (\ref{app:PiIV}).

Rewriting the above expressions in terms of the derivatives of energy
and following the procedure discussed in Appendix
(\ref{app:Leibnitz}), the contribution to the density of states by the
fourth-order WKB modes is given by:
{
\begin{eqnarray}
\Gamma_{4} (E) &=& \frac{\hbar}{\pi}\int\limits_{\rHo +h}^{L}dr
\left[-4 P^{(0)}_{4}(r)
\frac{\partial^2}{\partial \cE^2} \int\limits_{0}^{\cE} d\lambda 
\, \CGE \r. \nn \\
& & \qquad ~~ + \, \frac{8}{3}P^{(1)}_{4}(r) \frac{\partial^3}{\partial
\cE^3}\int\limits_{0}^{\cE}d\lambda \, \lambda \, \CGE  \nn \\
%%%%%%%
& &  \qquad ~~ - \, \frac{16}{15}P^{(2)}_{4}(r)
\frac{\partial^4}{\partial \cE^4}\int\limits_{0}^{\cE}
d\lambda \, \lambda^2 \, \CGE  \\
&& \qquad ~~ + \,  \frac{32}{105}P^{(3)}_{4}(r) \frac{\partial^5}{\partial
\cE^5}\int\limits_{0}^{\cE} d\lambda \, \lambda^3 \, \CGE  \nn \\
%%%%%%%
&& \qquad ~~ - \, \l. \frac{64}{945} P^{(4)}_{4} (r) \frac{\partial^6}{\partial
\cE^6}\int\limits_{0}^{\cE}d\lambda \, \lambda^4 \,\CGE\right]\, . \nn 
\end{eqnarray}
}
Integrating over $\lambda$, we get 
\beq
\Gamma_4 (E) = \frac{c_0^{(4)}}{E}  
\int\limits_{\rHo +h}^{L}dr \Sigma^{(4)}(r) \, , 
\label{eq:Gamma-4}
\eeq
where $c_0^{(4)}$ is a constant and $\Sigma^{(4)}(r)$ is given
in Eq. (\ref{eq:SigIVorder}). Using the expansion (\ref{eq:fg-serh})
close to the horizon, we get
{\small
\beq
\label{eq:Gamma4Fin}
\Gamma_4 (E) = \frac{c_0^{(4)} \kappa}{E} \l[
\frac{323 \, \rHo \kappa }{2520 \, (r-\rHo)^2} 
+ \frac{5 \, \rHo^2 \, g''(r) - 20 \, \kappa \rHo }{16 (r-\rHo)}\r] \, .
\eeq
This is the second result of the paper, regarding which we would like
to stress the following points: (i) The fourth order contributions to
the density of states goes as $1/E$. Using the expression
(\ref{eq:def-F}), it is easy to see that the fourth order contribution
to the free energy is independent of $\beta$ and, hence, the
contribution to the entropy vanishes\footnote{Note that, as mentioned
earlier, the free-energy integral has an infra-red ($E \to 0$)
divergence.}.
(ii) The density of states contribution close to the horizon again
depends only on the first and second order derivatives of the metric.
(ii) Comparing the fourth order contribution to the density of states
with the leading and second order, it is clear that the density of
states scales as $E^{3 - 2 n}$ where $n$ is the order of the WKB
modes.

%%%%%%%%%%%%%%%%%%%%%%%%%%%%%%%%%%%%%%%%%%%%%%%%%%%%%%%%%%%%%%%%%%%%%%%%%%%%%%%
\subsection{Sixth order}\label{sec:4D-VIorder}
Using the expression (\ref{eq:for-P6}), we get,
%
%{
\br
\label{eq:p6-f=g}
P_6(r) &=& \frac{P^{(0)}_6 (r)}{\CGEV{5}} 
+ \frac{\lambda(r) P^{(1)}_{6}(r)}{\CGEV{7}}  
%%%%%%
+ \frac{\lambda^{2}(r) P^{(2)}_{6}(r)}{\CGEV{9}} \nn \\
&+& \frac{\lambda^{3}(r) P^{(3)}_{6}(r)}{\CGEV{11}} 
+ \frac{\lambda^{4}(r) P^{(4)}_{6}(r)}{\CGEV{13}} 
+ \frac{\lambda^{5}(r) P^{(5)}_{6}(r)}{\CGEV{15}} \nn \\
& +&  \frac{\lambda^{6}(r) P^{(6)}_{6}(r)}{\CGEV{17}} 
\er
%}
%
where, the complete form of $P^{(i)}_{6}(r)$ [where $i = 0 \cdots 6$]
are given in Appendix (\ref{app:PiVI}).

Rewriting the above expressions in terms of the derivatives of energy
and following the procedure discussed in Appendix
(\ref{app:Leibnitz}), the contribution to the density of states by the
sixth-order WKB modes is given by:
\begin{eqnarray}
\Gamma_{6} (E) &=& 
\frac{\hbar^3}{ \pi}\int\limits_{_{\rHo +h}}^{L} dr \left[
\frac{8}{3} P^{(0)}_{6}(r) \frac{\partial^3}{\partial \cE^3}
\int\limits_{0}^{\cE}d\lambda \, \CGE \r. \nn \\
& & - \frac{16}{15}P^{(1)}_{6}(r) \frac{\partial^4}{\partial \cE^4}
\int\limits_{0}^{\cE}d\lambda \, \lambda \, \CGE  \nonumber \\
%%%%%
& & + \frac{32}{105}P^{(2)}_{6}(r) \frac{\partial^5}{\partial \cE^5}
\int\limits_{0}^{\cE}d\lambda \, \lambda^2 \, \CGE \nn \\
& & -\frac{64}{945}P^{(3)}_{6}(r) \frac{\partial^6}{\partial \cE^6}
\int\limits_{0}^{\cE}d\lambda \, \lambda^3  \, \CGE  \\
%%%%% 
&& + \frac{128}{10395} P^{(4)}_{6} (r) \frac{\partial^7}{\partial \cE^7}
\int\limits_{0}^{\cE}d\lambda \, \lambda^4  \, \CGE \nn \\
& & - \frac{256}{135135} P^{(5)}_{6} (r) \frac{\partial^8}{\partial \cE^8}
\int\limits_{0}^{\cE}d\lambda \, \lambda^5  \, \CGE \nonumber \\
%%%%%
& & + \left. 
\frac{512}{2027025} P^{(6)}_{6} (r) \frac{\partial^9}{\partial \cE^9}
\int\limits_{0}^{\cE}d\lambda \, \lambda^6  \, \CGE\right], \nonumber
\label{gamma6}
\end{eqnarray}
%\end{widetext}
%
%
Integrating over $\lambda$, we get 
\beq
\label{eq:gamma6fin}
\Gamma_6 (E) = \frac{c_0^{(6)}}{E^3}  \int\limits_{\rHo +h}^{L}dr
\Sigma^{(6)}(r)
\eeq
where $c_0^{(6)}$ is a constant $\Sigma^{(6)}(r)$ is given by
Eq. (\ref{eq:SigVIorder}). Repeating the steps i. e. using the
relation (\eq{eq:def-F}) obtaining the free-energy, substituting the
free-energy in (\ref{eq:CE-entropy}) and expanding the metric close to
horizon using Eq. (\ref{eq:fg-serh}), we get,
\begin{eqnarray}
\frac{S_6}{\epsilon} &=& -\frac{13892\pi^2}{45045}\frac{\rHo^2}{h_c^2} \nn \\
&+& \l[\frac{9 \pi^2}{77}  g''(\rHo) \rHo^2  + \frac{30 \pi^2}{77} 
\kappa \, \rHo \r] \log\l(\frac{\rHo^2}{h_c^2}\r) \, .
\label{en6I}
\end{eqnarray}
This is the third result of the paper, regarding which we would like
to stress the following points:
(i) The sub-leading corrections (like the zeroth and second order WKB)
depend only on the surface gravity and second derivative of the metric
functions.  This indeed implies that brick-wall entropy does indeed
provide generic corrections to the Bekenstein-Hawking entropy at all
orders. We have shown this to be the case upto sixth order. It is
natural to expect this to be valid for all higher orders.
(ii) As mentioned above, the density of states in each order is given
by $E^{3 - n}$.
(iii) $\epsilon$ in the above expression is due to the fact that
the $E$ divergences as $E \to 0$. Thus, the above expression for 
the entropy depends on the infra-red cutoff.

%%%%%%%%%%%%%%%%%%%%%%%%%%%%%%%%%%%%%%%%%%%%%%%%%%%%%%%%%%%%%%%%%%%%%%%%%%%%%%%

\section{Higher order contributions in six dimensions}
\label{sec:HD-BW}

In this section, we obtain the zeroth and second-order WKB mode 
contributions to the brick-wall
entropy in six-dimensional black hole spacetime. The analysis can be
extended to any even dimensional spacetime, however, the analysis in
odd-dimensional spacetime is more involved\footnote{This can be
traced to the fact that the wave propagation in these spacetimes are
non-local. For more discussion, see Refs.
\cite{Barton:1989,Courant-Hilb:1962b}.}.

We show that the results of the brick-wall entropy in the zeroth and
second order WKB modes have the same structure confirming the results
of 4-dimensions and has the following generic form:
\beq
\SBW^{\rm (6D)} = \SBH + \GAH + \FAHSD \,
\log\l(\frac{\AHo}{\lPl^2}\r) \, .
\eeq

\subsection{Zeroth order}
\label{sec:0order-6d}

In the case of $D = 4$, the weight function (\ref{eq:Weightl}) becomes
\beq
\label{eq:weight-6D}
{\cal W}(\ell) = \frac{(\ell + 1) (\ell +  2)}{6} \, .
\eeq
Substituting the above expression in (\ref{eq:Gamma-order}), the
density of states for the zeroth-order WKB modes (for $f(r) = g(r)$)
is
\begin{eqnarray}
\Gamma_{0}^{\rm (6D)} =
\frac{1}{\hbar^3\pi} \int\limits_{_{r_{_{H}}+h}}^{L}
dr \frac{r^{2}}{g(r)^2} 
\int\limits_{0}^{{\cal E}} d\lambda \left(\frac{\lambda r^2 }{\hbar^2
g(r)} +2 \right)  \CGE
\end{eqnarray}
where, $\CGE$ is given by (\ref{eq:def-CGE}) and 
\beq
\lambda =  l(l+3) \hbar^2 \frac{g(r)}{r^2} \, .
\eeq
Repeating the procedure discussed in the previous section, the zeroth
order brick-wall entropy is given by
\begin{eqnarray}
S_{0}^{\rm (6D)} = 
\frac{32 \pi^5 }{945 \beta^5 \hbar^5}
\int\limits_{\rHo+h}^{L} \!\!\! dr \frac{r^4}{g(r)^3} +
\frac{8 \pi^2 }{135 \beta^3 \hbar^3}
\int\limits_{\rHo+h}^{L} \!\!\! dr \frac{r^2}{g(r)^2} \, .  
\end{eqnarray}
Expanding the metric near the horizon (\ref{eq:fg-serh}), upto third order, 
and using the relation (\ref{eq:defhc-rind}), the zeroth-order entropy 
is given by
\br
\label{entropy6DL}
S_{0}^{\rm (6D)} &=& \frac{\rHo^{4}}{3780 h_{c}^{4}} 
+ \GAHOz + \FAHOz \log{\left(\frac{r_{H}^{2}}{h^{2}_c }\right)} \, ,
\er
where 
\br
\GAHOz & =& \frac{\rHo^2}{15120 \, h_c^2} 
\left[- 3 g''(\rHo) \, \rHo^2 +16 \kappa  \rHo + 56\right] \nn \\
%%%
\FAHOz &=& + \frac{\rHo}{60480}
\l[\l( 2 \kappa g'''(\rHo) - 3 g''(\rHo)^2\r) \rHo^3 \r.  \\
&+& 24  g''(\rHo) \kappa \rHo^2 - 224 \kappa   
 + \l. \l(56  g''(\rHo) -48 \kappa ^2\r) \rHo 
\r] \nn \, .
\er
As in the 4-dimensions, the leading order term in the above expression
(\ref{entropy6DL}) is proportional to area (Bekenstein-Hawking area
relation).  The sub-leading term has two parts: (a) One that contain a
purely power-law corrections [$\GAHOz$] that is absent in the
case of 4-dimensions.  (b) The logarithmic term contains a prefactor
which, in general, is a function of area as like in
4-dimensions. $\FAHOz$, as in 4-dimensions, depend upto the second
derivative of the metric close to the horizon while $\GAHOz$ depend
upto the third derivative of the metric close to the horizon.

\subsection{Second order}
\label{sec:2order-6d}

Substituting Eq. (\ref{eq:weight-6D}) in Eq. (\ref{eq:for-P2}), the
contribution to the density of states from the second-order WKB modes
is given by
\begin{eqnarray}
\Gamma_2^{(6D)} &=& \frac{1}{\pi \hbar}
\int\limits_{\rHo +h}^{L} dr 
\left[ P_{26}^{(0)}(r) \int\limits_{0}^{{\cal E}} d\lambda  
{\cal W}(\lambda) \frac{\partial \CGE}{\partial {\cal E}}\r. \nn \\
%%%%
& + & \l. 
4 P_{26}^{(1)}(r) \int\limits_{0}^{{\cal E}} d\lambda \, {\cal
W}(\lambda) \, \lambda(r) \,  
\frac{\partial^2 \CGE}{\partial {\cal E}^2} \right. \\  \nonumber 
%%%%%
&+& \left. \frac{8}{3} P_{26}^{(2)}(r)
\int\limits_{0}^{{\cal E}} d\lambda \, {\cal W}(\lambda) \, \lambda(r)^2 \,
\frac{\partial^3 \CGE}{\partial {\cal E}^3}  \right] \, , 
\end{eqnarray}
where, the degeneracy factor ${\cal W}(\lambda)$ in terms of $\lambda$
is given by
\begin{eqnarray}
{\cal W}(\lambda) = \frac{\lambda}{\hbar^2} \frac{r^2 }{g(r)}+2 \, ,
\end{eqnarray}
and 
\begin{eqnarray}
P_{26}^{(0)}(r) &=& -\frac{g(r)}{r^2} - \frac{g'(r)}{r} \, , \\ \nonumber
P_{26}^{(1)}(r) &=& \frac{3 g(r)^2}{4 r^4}-\frac{3g(r) g'(r)}{4
r^3}+\frac{g'(r)^2}{8 r^2}+\frac{ g(r) g''(r)}{8 r^2} \, , \\ \nonumber
P_{26}^{(2)}(r) &=&\frac{5g(r)^3}{8 r^6}-\frac{5g(r)^2 g'(r)}{8 r^5}+\frac{5
g(r) g'(r)^2}{32 r^4} \, .
\end{eqnarray}
Repeating the procedure discussed in Appendix (\ref{app:Leibnitz}), and
substituting the resultant in Eq. (\ref{eq:def-F}), we get,
{\small
\begin{eqnarray}
&& \!\!\!\!\!\!\!\!
F_{_{2}} = \frac{\pi}{\beta^2 \hbar} \!\!\!
\int\limits_{r_{H} +h}^{L} \!\!\! dr 
\l[-\frac{r^2 g'(r)^2}{54 g(r)^2}
+\frac{3 g''(r) r^2 +14 g'(r) r}{108 g(r)}
+\frac{5}{54}\r] \\
%%%%
&& \!\!\!\!
+ \frac{\pi^3}{\beta^4 \hbar^3} \!\!\!
\int\limits_{r_{H} +h}^{L} \!\!\! dr \l[
\frac{g''(r) r^4+6 g'(r) r^3}{135 g(r)^2}
- \frac{r^4 g'(r)^2}{90 g(r)^3} 
- \frac{2 r^2}{135 g(r)}\r] \, \, . \, \nn
\end{eqnarray}
}
Substituting the above expression in Eq. (\ref{eq:CE-entropy}) and
expanding the metric using (\ref{eq:fg-serh}) the second order WKB
mode contribution to the brick-wall entropy is given by
\br
S_{2}^{\rm (6D)} &=& \frac{r_{H}^4}{180 h_{c}^4} 
+ \GAHSz + \FAHSz \log{\left(\frac{r_{H}^{2}}{h^{2}_c }\right)} \, ,
\label{entropy6D2}
\er
where 
\br
\GAHSz & =& \frac{\rHo^2}{2160 \, h_c^2} \left[80 - \rHo^2 g''(\rHo) \right]
\nonumber  \\
%%%%%
\FAHSz &=& - \frac{\rHo}{8640} \l[ 
\l(g''(\rHo)^2 - 2 \kappa g'''(\rHo)\r) \rHo^3 \r.  \\
& & \l. \!\!\!\!\! - 8 g''(\rHo) \kappa \rHo^2 - 800 \kappa  
- (112 \kappa^2  + 40 g''(\rHo)) \rHo  \r] \, . \nn 
\er
This is the fourth result of this paper, regarding which we would like
to discuss the following: (i) As in the case of 4-dimensions, the
second order WKB modes contribute the same to the brick-wall entropy
as the zeroth order modes. This again proves that in-order to
associate the brick-wall entropy to the black hole entropy, it is
necessary to calculate all the higher order WKB mode
contribution. (ii) As in the case of 4-dimensions, the dependence of
the entropy on the horizon area is the same in both the orders. (iii)
$\GAHSz$ (like $\GAHOz$) has a generic power-law corrections to $\SBH$
and depend only upto the second derivative of the metric near the
horizon. $\FAHSz$ (like $\FAHOz$) has a prefactor which is a function
of the area, and --- as in 4-dimensions -- is a constant only for the
Schwarzschild spacetime.

%%%%%%%%%%%%%%%%%%%%%%%%%%%%%%%%%%%%%%%%%%%%%%%%%%%%%%%%%%%%%%%%%%%%%%%%%%%%%%%

\section{Results for specific black holes}
\label{sec:appl-BH}

In this section, we shall explicitly write down the brick wall entropy 
(evaluated upto the second order in the WKB approximation) for a few 
well-known black hole solutions in four and six spacetime dimensions.
We shall restrict ourselves to the cases wherein $f(r)=g(r)$. 

%%%%%%%%%%%%%%%%%%%%%%%%%%%%%%%%%%%%%%%%%%%%%%%%%%%%%%%%%%%%%%%%%%%%%%%%%%%%%%%

\subsection{Four dimensional examples}

We find that, in four dimensions, on combining the zeroth 
order~(\ref{eq:EntrF}) and the second order~(\ref{en2I}) 
terms, the total brick wall entropy can be expressed as
\beq
\SBW^{\rm (4D)} 
= \SBH + \FAHFD\; \log\l(\frac{\AHo}{\lPl^2} \r) ,
\label{eq:4D-Tot-SBW}
\eeq
where, in order for the leading term to match the Bekenstein-Hawking
entropy~(\ref{eq:BHentropy}), we have set the brick wall invariant
cutoff $h_{c}$ to be 
\beq
h_{c}^2 = \l(\frac{11\; \lPl^2}{90 \pi}\r)\,.
\label{eq:4dcutoff}
\eeq
and the quantity $\FAHFD$ is given by
\beq
\label{eq:defFAHFD}
\FAHFD = - \l(\frac{1}{60}\r)\, g''(\rHo)\; \rHo^2 
- \l(\frac{1}{10}\r)\, \kappa\, \rHo\,.
\eeq
%

%%%%%%%%%%%%%%%%%%%%%%%%%%%%%%%%%%%%%%%%%%%%%%%%%%%%%%%%%%%%%%%%%%%%%%%%%%%%%%%

\subsubsection{Schwarzschild black hole}

For the Schwarzschild black hole, the metric coefficients are given 
by Eq.~(\ref{eq:4DSch}) and the event-horizon of the black hole is 
located at $r_{_{\rm H}} = (2 M)$.  
The surface gravity $\kappa$ and the second derivative of the metric 
at the horizon are given by
\beq
\kappa = \l(\frac{1}{4 M}\r),\quad 
g''\l(r_{_{\rm H}}\r) = -\l(\frac{1}{2 M^2}\r)\,.
\eeq
On substituting these expressions in \eq{eq:4D-Tot-SBW}, we obtain 
that 
\beq
S_{\rm Sch}^{\rm (4D)} 
= \SBH 
- \l(\frac{1}{60}\r)\, \log{\left(\frac{\AHo}{\lPl^2}\right)}.
\label{eq:schfr}
\eeq
%

%%%%%%%%%%%%%%%%%%%%%%%%%%%%%%%%%%%%%%%%%%%%%%%%%%%%%%%%%%%%%%%%%%%%%%%%%%%%%%%

\subsubsection{Schwarzschild (anti-)de sitter spacetime}

For the Schwarzschild (anti-)de sitter spacetime, the metric 
function $g(r)$  is given by 
\beq
g(r) = \l(1 - \frac{2 M}{\tilde r} \pm \frac{{\tilde r}^2}{l^2}\r) 
= \l(1 - \frac{2}{r} \pm \frac{r^2}{y}\r)\label{eq:gradss}
\eeq
where $y = (l/M)^2$, $r =({\tilde r}/M)$, $M$ is the mass of the
black hole, $l$ is related to the positive (negative) cosmological 
constant and $-(+)$ corresponds to (anti-)de Sitter spacetime.
Note that the coordinates $y$ and $r$ are dimensionless. 
While the Schwarzschild anti-de Sitter spacetime has only one horizon 
associated with the singularity at the origin, the Schwarzschild 
de Sitter has two---one event and one cosmological--- horizons. 
Here, we shall focus on the entropy associated with the event-horizon. 

Recall that the event horizon is identified by the condition 
$g(r)=0$.  
On substituting the resulting $\rHo$ corresponding to the above 
$g(r)$ in \eq{eq:4D-Tot-SBW}, we find that the brick wall entropy
upto the second order can be expressed as
\begin{eqnarray}
\!\!
S_{\rm Sch-(a)dS}^{\rm (4D)} 
= \SBH - \l(\frac{\pi^{1/2}}{15\, \AHo^{1/2}} 
+ \frac{\AHo}{\pi\, y}\r)\;
\log{\left(\frac{M^2\, \AHo}{\lPl^2}  \right)} \, , 
\end{eqnarray}
where $\AHo$ defined in-terms of the coordinate $r$ is also dimensionless.
In contrast to the purely Schwarzschild case  wherein the prefactor 
to the logarithmic correction was a constant, here the factor is a 
function of the horizon area. 

%%%%%%%%%%%%%%%%%%%%%%%%%%%%%%%%%%%%%%%%%%%%%%%%%%%%%%%%%%%%%%%%%%%%%%%%%%%%%%%

\subsubsection{Reissner-Nordstr\"om black hole}

For the Reissner-Nordstr\"om black hole, we have 
\beq
g(r) = \l(1 - \frac{2M}{\tilde r} + \frac{Q^2}{{\tilde r}^2}\r)
= \l(\frac{(r - r_{-}) (r - r_+)}{r^2}\r)\,,
\eeq
where $M$ and $Q$ denotes the mass and the electric charge of the black
hole.
Also, $r = {\tilde r}/M$ and $r_{\pm}$ is the outer/inner horizon given 
by 
\beq
r_{\pm} = \l(1 \pm \sqrt{1 -\frac{Q^2}{M^2}}\r)\, ,
\eeq 
where, again, $r$ is a dimensionless variable.
It is the outer horizon $r_{+}$ that is the event horizon of the black
hole.

On substituting the above relations in \eq{eq:4D-Tot-SBW}, we obtain
the brick wall entropy upto the second order to be
\beq
S_{\rm RN}^{\rm (4D)} 
= \SBH 
- \l(\frac{\pi^{1/2}}{15 \, \AHo^{1/2}}\r)\; 
\log{\left(\frac{M^2\,\AHo}{\lPl^2} \right)}\, ,
\label{eq:rnfr}
\eeq
where, again, $\AHo$ defined in-terms of $r$ is dimensionless.
As in the previous example, the prefactor again turns out to be a
function of the horizon area $\AHo$. 

%%%%%%%%%%%%%%%%%%%%%%%%%%%%%%%%%%%%%%%%%%%%%%%%%%%%%%%%%%%%%%%%%%%%%%%%%%%%%%%

\subsection{Six dimensional examples}

On combining the zeroth order~(\ref{entropy6DL}) and the second 
order~(\ref{entropy6D2}) terms, we find that the brick wall 
entropy for six dimensional black holes can be expressed as
\beq
\SBW^{\rm (6D)} 
= \SBH + \GAH + \FAHSD\, \log\l(\frac{\AHo}{\lPl^4}\r) ,
\label{eq:6D-Tot-SBW}
\eeq
where, as in the four dimensional case, we have chosen the invariant 
cutoff $h_{c}$ to be such that the leading term matches the
Bekenstein-Hawking entropy.
The quantities $h_{c}$, $\GAH$, $\FAHSD$ are given by
\br
\!\!\!\!\!\!\!\!\!\!\!\!
\label{eq:6dcutoff}
h_{c}^4\! &=&\! \l(\frac{11\, \lPl^4}{1860 \pi^2}\r)\,,\\
\!\!\!\!\!\!\!\!\!\!\!\!
\FAHSD\! &=&\!  \l(\frac{\rHo}{30240}\r)\, 
\l[\l(8\, \kappa\, g'''(\rHo)-5\, {g''}^{2}(\rHo)\r)\,\rHo^3\r.\nn\\
& & \qquad\qquad\;\;
+\; 40\,  g''(\rHo)\, \kappa\, \rHo^2 + 2688\, \kappa\nn\\
& &\qquad\qquad\;\;
+\; \l.8\, \l(46\, \kappa^2 + 21\, g''(\rHo)\r)\, \rHo \r]\,,\\
\!\!\!\!\!\!\!\!\!\!\!\!
\GAH 
\!&=&\! \l(\frac{31\, \rHo^4}{165\, \lPl^4}\r)^{1/2}\, 
\l(\frac{\pi}{252}\r)\nn\\
& &\qquad\quad\times\; \l(-5\, g''(\rHo)\, \rHo^2  
+ 8\, \kappa\, \rHo + 308\r)\,. 
\er

%%%%%%%%%%%%%%%%%%%%%%%%%%%%%%%%%%%%%%%%%%%%%%%%%%%%%%%%%%%%%%%%%%%%%%%%%%%%%%%

\subsubsection{Schwarzschild black hole}

In six dimensions, the function $g(r)$ for Schwarzschild black holes
is given by
\beq
g(r) = \l[1 - \left(\frac{r_{_{0}}}{r}\right)^{3}\r]\,,
\eeq
where $r_{_{0}}$ is related to the black hole mass $(M)$ by the relation 
\beq
M = \l(\frac{2\, \pi\, r_{_{0}}^{3} }{3 \pi G_{_{6}}}\r)\,,
\eeq
with $G_{6}$ being the six-dimensional Newton's constant (which we shall
hereafter set to unity). 
On using the definition ~(\ref{eq:kappa-def}) of the surface gravity 
$\kappa$ we find that
\beq
\kappa = \l(\frac{3}{2 \, \rHo}\r)\,,
\eeq
where $\rHo = r_{_{0}}$. 
Substituting the derivatives of the above metric function $g(r)$ in 
the expression~\eq{eq:6D-Tot-SBW}, we obtain the brick wall entropy to
be 
\begin{eqnarray}
\!\!\!\! S_{\rm Sch}^{\rm (6D)} 
&=& \SBH 
+ \frac{19 \pi}{63} \sqrt{\frac{155}{33}} \frac{\rHo^2}{\lPl^2}
- \frac{59}{840} \log\left(\frac{\AHo}{\lPl^4}\right) \nn \\
%%%%% 
&=& \SBH 
+ \frac{19}{63} \sqrt{\frac{155}{88}} \frac{\AHo^{1/2}}{\lPl^2}
- \frac{59}{840} \log\left(\frac{\AHo}{\lPl^4}\right) 
\end{eqnarray}
where in deriving the above expression we have used the expression 
for the area of the 4-sphere i.e. $\AHo = (8 \pi^2)/3 \rHo^4$. 

Unlike four-dimension, there is a pure power-law correction term 
to the Bekenstein-Hawking entropy.  

%%%%%%%%%%%%%%%%%%%%%%%%%%%%%%%%%%%%%%%%%%%%%%%%%%%%%%%%%%%%%%%%%%%%%%%%%%%%%%%

\subsubsection{Schwarzschild (Anti-)De sitter black hole}

The line-element for Schwarzschild (Anti-)De sitter 
spacetime is given by (\ref{eq:spher-tr}) with 
\beq
f(r) = g(r) = 1 - \l(\frac{r_{_{0}}}{\tilde r}\r)^3 \pm 
\frac{{\tilde r}^2}{l^2} = 1 - \frac{1}{r^3} \pm \frac{r^2}{y_{_{6}}}
\eeq
where $y_{_6} = (l/r_{_{0}})^2$, 
$r \to ({\tilde r}/r_{_{0}})$, $l$ is related to the
positive (negative) cosmological constant and 
$-(+)$ corresponds to asymptotic (Anti-)de Sitter. Here again,
for Schwarzschild-de Sitter, we consider only the event-horizon.

The event horizon is given by the condition $g(r=\rHo)=0$.  
Substituting these in Eq. (\ref{eq:6D-Tot-SBW}), we get,
{\small
\begin{eqnarray}
& & \!\!\!\!\!\!\!\!\! S_{\rm (A)dS}^{\rm (6D)} = \SBH 
%%%%
- \frac{\pi}{252} \sqrt{\frac{31}{165}}\frac{r_{_{0}}^2}{\lPl^2} 
\l[ \frac{2 \rHo^4}{y_{_6}} - 308 \rHo^2 -\frac{72}{\rHo} \r]  \\
%%%%%%
&-& \!\!\! \l[\frac{19}{1512}\frac{\rHo^4}{y_{_6}^2} 
+ \frac{\rHo^2}{10 y_{_6}} 
+ \frac{11}{252 y_{_6} \rHo}
+ \frac{1}{15 \rHo^3}
\r] \log\left[\frac{r_{_{0}}^4}{\lPl^4} \AHo\right] \, . \nn
\end{eqnarray}
}
Note that as in four-dimensions, $\AHo$ is dimensionless.
The above expression gives a series of power-law corrections to $\SBH$. 

%%%%%%%%%%%%%%%%%%%%%%%%%%%%%%%%%%%%%%%%%%%%%%%%%%%%%%%%%%%%%%%%%%%%%%%%%%%%%%%

\subsubsection{Reissner Nordstr\"om black hole}

For 6-dimensional Reissner Nordstr\"om black hole, we have 
\beq
f(r) = g(r) 
= 1 - \l(\frac{r_{_{0}}}{\tilde r}\r)^3 + \frac{\theta^2}{{\tilde r}^6} 
= 1 - \l(\frac{1}{\tilde r}\r)^3 + \frac{\chi}{r^6}
\eeq
where the charge of the black hole is given by
\br
Q =  \frac{3}{2 \pi} \frac{\theta^2}{G_{_6}} \, ; 
\, r = \frac{\tilde r}{r_{_{0}}} \, ; \, \chi = \frac{\theta^2}{r_{_{0}}} \, .
\er
As in 4-dimensions, this has two horizons -- event and Cauchy horizon. 
The event horizon $(\rHo)$ is the outer horizon while the inner 
horizon is the Cauchy horizon. Note that $\chi$ and $r$ are dimensionless.

Substituting these in Eq. (\ref{eq:6D-Tot-SBW}), we get,
{\small
\br
& & S_{\rm RN}^{\rm (6D)} = \SBH 
%%%%
+ \frac{\pi}{252}\sqrt{\frac{31}{165}}\frac{r_{_{0}}^2}{\lPl^2} 
\l[ 308 \rHo^2 + \frac{72}{\rHo}- \frac{234 \chi}{\rHo^4}\r] \\
%%%%%
& & + \l[-\frac{23 \chi ^2}{280 \rHo^{12}} 
+ \frac{\chi }{140 \rHo^9} + \frac{3 - 28 \chi}{840 \rHo^6}
+ \frac{1}{15 \rHo^3} \r]\log\left[\frac{r_{_{0}}^4}{\lPl^4}\AHo\right] \, , 
\nn
\er
}
where, again, $\AHo$ is dimensionless. As can be seen, this also 
generates a series of power-law corrections to the Bekenstein-Hawking 
entropy.

%%%%%%%%%%%%%%%%%%%%%%%%%%%%%%%%%%%%%%%%%%%%%%%%%%%%%%%%%%%%%%%%%%%%%%%%%%%%%%%

\subsubsection{Boulware-Deser black hole}

Boulware-Deser black-hole \cite{Boulware-Dese:1985a} is an exact  
spherically symmetric solution of the Einstein action modified by 
the quadratic Gauss-Bonnet combination i.~e.
\beq
I = \int d^6 x \sqrt{-g} \l[R + \alpha_{_{\rm gb}}
\l(R^2 - 4 R_{ab} R^{ab} + R_{abcd} R^{abcd} \r) \r]
\eeq
where $\alpha_{_{\rm gb}}$ is the Gauss-Bonnet coupling. The line-element 
is given by (\ref{eq:spher-tr}) where
\br
\label{eq:6DBD}
f(r) = g(r)&=& 
1 + \frac{{\tilde r}^2}{6 {\tilde \lambda}} 
\left(1 - \sqrt{1 +\frac{12 \omega {\tilde \lambda}}{{\tilde r}^5}}
\right) \\
&=& 1 + \frac{r^2}{6 \lambda} 
\left(1 - \sqrt{1 +\frac{12 \lambda}{r^5}} \right) \nn 
\er
where ${\tilde \lambda} = 6 \alpha_{_{\rm gb}}$, $r = {\tilde
r} \, \omega^{-1/3}$, $\lambda = {\tilde \lambda} \, \omega^{- 2/3}$ and 
$\omega$ is related to the ADM mass $(M_{_{\rm ADM}})$ by the relation
\begin{equation}
M_{_{\rm ADM}} = \frac{\omega}{4 \pi} \, \AHo \, .
\end{equation}
Note that the rescaled variables $r, \lambda$ are dimensionless. The 
horizon is given by the condition $f(r) = 0$ and occurs at $r = \rHo$ 
such that 
\beq
\rHo^{3} + 3 \lambda \rHo -1 = 0 \, .
\eeq
The existence of the horizon requires $\lambda > 0$ and which then 
gives $rH^{3} < 1$.
The surface gravity of the event horizon is given by
\beq
\kappa = \frac{\omega^{-1/3}}{2 \rHo} 
\l(\frac{1+ 2 \rHo^{3}}{2 - \rHo^{3}}\r) \, .
\eeq
Substituting these in Eq. (\ref{eq:6D-Tot-SBW}), we get,
{\scriptsize
\br
S_{\rm GB}^{\rm (6D)} &=& \SBH  + \sqrt{\frac{148955}{523908}} 
\frac{\omega ^{2/3}\pi}{\lPl^2} 
%\\ & & \qquad \times 
\l[\rHo^2  + \frac{7  \rHo^2}{31 (2 - \rHo^3)} - 
\frac{25  \rHo^2}{31 (2 - \rHo^3)^2}\r] \nn \\
%%%%%
&+& \left[\frac{125}{216}\frac{1}{(2-\rHo^3)^6}
-\frac{1175}{756}\frac{1}{(2-\rHo^3)^5} + 
\frac{725}{504}\frac{1}{(2-\rHo^3)^4}
- \frac{649}{7560}  
\right. \nonumber \\ 
%%%%%
&+& \left.
\frac{355}{756}\frac{1}{(2-\rHo^3)^2} + \frac{17}{189}\frac{1}{2-\rHo^3}
- \frac{655}{756} \frac{1}{(2-\rHo^3)^3} \right] 
\log\left[\frac{\omega^{4/3}}{\lPl^4}\AHo\right] \nn \\ 
\label{eq:BDentropyBW}
\er
}
It is instructive to stress the implications of this result: (a) In
the limit of $\lambda \to 0$ limit $S_{\rm GB}^{\rm (6D)}$ reduces to
$S_{\rm Sch}^{\rm (6D)}$ . (ii) The subleading corrections to the
Bekenstein-Hawking entropy for the Boulware-Deser black-hole have 
been obtained earlier by other authors (see, for instance, Ref. 
\cite{Myers-Simo:1988a}). Using the Noether charge, it was shown that
\beq
S_{\rm GB}^{\rm NC} = \SBH 
+ \frac{8 \pi^2\, \omega^{-4/3}}{\lPl^{4}}  \lambda \rHo^2
\eeq
Comparing the two results we see that the subleading corrections
in the brick-wall approach, unlike the Noether charge approach, can be
completely specified by the horizon properties.

%%%%%%%%%%%%%%%%%%%%%%%%%%%%%%%%%%%%%%%%%%%%%%%%%%%%%%%%%%%%%%%%%%%%%%%%%%%%%%%

\section{Discussion}\label{sec:discussion}

\subsection{Summary}

As we have pointed out repeatedly, the brick wall model has 
been a very popular approach that has been utilized to recover 
the Bekenstein-Hawking entropy $\SBH$ in a multitude of 
situations~\cite{Paddy:1986a,Mann-Tara:1990a,Ghosh-Mitr:1994a,
Demers-Lafr:1995,Frolov-Furs:1995a,Kim-Kim:1996a,Kim-Kim:1996bp,
Solodukhin:1996a,Shen-Chen:1997a,Cognola-Lecc:1997a,Ho-Won:1997a,
Solodukhin:1997a,Frolov-Furs:1998a,Ho-Kang:1998,Mukohyama-Isra:1998,
Winstanley:2000a}.
In all these efforts, it is only the leading term in the 
WKB expansion~(\ref{eq:Gamma}) that has been taken into 
account in evaluating the density of states and the 
associated free energy and entropy of quantum fields around 
black holes.
Also, the metric has almost always been assumed to be of 
the Rindler form near the event horizon.

In this work, we have extended the brick wall approach to the 
higher orders in the WKB approximation.
Moreover, by expanding the metric functions~$f(r)$ and~$g(r)$ 
beyond the leading order near the event horizon, we have been 
able to evaluate the corrections to the Bekenstein-Hawking entropy 
for spherically symmetric black holes in four and six dimensions.
To begin with, we have illustrated that, even the often considered 
zeroth order term in the WKB approximation leads to corrections to 
the Bekenstein-Hawking entropy, provided the metric functions are 
expanded beyond the linear order near the horizon.
Secondly, we have shown that all the higher order terms in the WKB
approximation have the same form as the zeroth order term.
Lastly, we find that, the higher order WKB terms actually contribute 
{\it more}\/ to the entropy than the lower order terms. 

Specifically, we have shown that, upto the second order in the 
WKB approximation, the brick wall entropy of four dimensional 
black holes can be expressed as
$$ 
\SBW^{\rm (4D)} 
= \SBH + \FAHFD\, \log\l(\frac{\AHo}{\lPl^2} \r)\,, 
$$
where $\FAHFD \propto \AHo^{n}$ with $n < 1$. 
Whereas, in six dimensions, we find that the brick wall entropy up 
to the second order has the form
$$ 
\SBW^{\rm (6D)} 
= \SBH + \GAH + \FAHSD\, \log\l(\frac{\AHo}{\lPl^2} \r)\,,
$$
where $\GAH \propto \AHo^{n}$ and  $\FAHSD \propto \AHo^{m}$ with
$(n, m)< 1$.
Note that, while the brick wall entropy in four dimensions depends
only on the first and the second derivatives of the metric at the 
horizon, in six dimensions, it depends on the third derivative as 
well. 
It is tempting to propose that, at least in even dimensions, the 
brick wall entropy will depend on as many as derivatives of the 
metric as half the number of spacetime dimensions!
However, the black hole entropy is a coordinate invariant concept.
If the brick wall entropy depends on arbitrary derivatives of the metric 
functions at the horizon, then it is not apriori evident that the 
resulting entropy will be coordinate invariant.
We believe that this is an issue that needs to be addressed
satisfactorily.

%%%%%%%%%%%%%%%%%%%%%%%%%%%%%%%%%%%%%%%%%%%%%%%%%%%%%%%%%%%%%%%%%%%%%%%%%%%%%%%

\subsection{Comparison with results from other approaches}

Power law and logarithmic corrections to the Bekenstein-Hawking 
entropy~$\SBH$ that we have obtained in the brick wall approach 
has been encountered earlier in a few other approaches to black 
hole entropy.
For instance, the Noether charge approach predicts a generic power 
law correction to the Bekenstein-Hawking entropy~\cite{Wald:1993a}.
However, unlike our approach wherein the brick wall entropy can
be completely expressed in terms of the metric and its first few 
derivatives at the event horizon, the Noether charge entropy can 
not be mapped to the horizon properties. 
It is also interesting to note that, in the case of the four 
dimensional Reissner-Nordstr\"om black hole, for large horizon 
area, i. e. when $M \gg \lPl$, the brick wall entropy 
$S_{\rm RN}^{\rm (4D)}$ [cf. \eq{eq:rnfr}] reduces to
\beq
S_{\rm RN}^{\rm (4D)} 
\simeq \SBH 
- \l(\frac{2 \pi^{1/2}}{15}\r)\; 
\l(\frac{1}{\AHo^{1/2}} - 
\frac{\lPl^2\, \AHo^{3/2}}{M^2}\r)\,.
\eeq
Similar power law corrections arise on evaluating the entanglement 
entropy of such black holes~\cite{Das-Shanki:2007b}. 
This behavior seem to suggest a possible relationship between the 
brick wall model and the approach due to entanglement entropy.
Another interesting feature is the absence of power law corrections 
in case of four dimensional Schwarzschild black hole. 
It seems to indicate that power law corrections to the Bekenstein-Hawking 
entropy are related with the presence of matter. 
The logarithmic corrections that we have obtained as 
in~\eq{eq:schfr} for the case of the four dimensional Schwarzschild 
black hole has also been arrived at in other methods such as the 
approach through conformal field theory~\cite{Carlip:2000a}, statistical 
fluctuations around thermal equilibrium~\cite{Das-Maju:2001a} and 
spin foam models~\cite{Kaul-Maju:2000a}. 
However, it should be pointed out that the prefactor to the logarithmic 
term that we obtain turns out to be different from the one that arises 
in the other approaches.

%%%%%%%%%%%%%%%%%%%%%%%%%%%%%%%%%%%%%%%%%%%%%%%%%%%%%%%%%%%%%%%%%%%%%%%%%%%%%%%

\subsection{A few words on the divergences}

The divergence that results in the need of a brick wall cut-off
arises even at the leading order in the WKB approximation, and 
is, obviously, well-known.
So, it is not all suprising that such a divergence occurs at the 
higher orders terms as well.
However, in addition to the brick wall divergence, we seem to
encounter three more types of divergences at the higher orders.
The first is the divergence that occurs at the upper limit when 
integrating over~$\ell$ in the higher orders and the second is 
an infra-red divergence that arises at a sufficiently high order 
when integrating over~$E$ (as in the case of the sixth order term
in four dimensions).
As we mentioned above, the higher order terms contribute more to 
the brick wall entropy than the lower orders.
Therefore, the third and last is a divergence that can arise if 
we get around to summing over all the terms~$n$.
 
The first of these additional divergences is associated with the 
turning points.
Such divergences are known to occur at the higher orders in the 
WKB approximation even in non-relativistic quantum mechanics.
Evidently, these divergences are not a feature of the field 
theory, but a feature of the approximation.
The procedure we have adopted to isolate and discard these 
divergences effectively deals with them.

In contrast, the remaining two divergences that occur at the 
higher orders are field theoretic divergences.
The infra-red divergence is clearly one as it arises when 
integrating over all the modes.
In the context of the leading order results, it has 
been argued that the brick wall divergence can be
absorbed into the renormalization of the Newton's 
constant~\cite{Demers-Lafr:1995,Mukohyama-Isra:1998}.
Clearly, if the higher order WKB terms continue to contribute
more to the brick wall entropy than the lower order ones, then
a divergence will arise when the contributions from all the 
orders are summed over.
We believe that, when working at the higher orders in the WKB
approximation, these additional divergences need to be accommodated 
in a renormalization procedure, along with the brick wall divergence 
itself.

%%%%%%%%%%%%%%%%%%%%%%%%%%%%%%%%%%%%%%%%%%%%%%%%%%%%%%%%%%%%%%%%%%%%%%%%%%%%%%%

\subsection{Outlook}

Since the odd dimensional cases are analytically more involved,
after first working in four dimensions, we had jumped to consider
six-dimensional black holes.
Needless to add, it will be interesting to extend the current 
analysis to black holes in odd dimensional spacetimes. 
The Banados-Teitelboim-Zanelli black hole in three 
dimensions~\cite{BTZ:1992,BTZ:1993} and the five-dimensional 
Boulware-Deser black hole \cite{Boulware-Dese:1985a} are 
interesting cases that are to be studied.
The canonical entropy has been calculated {\it exactly}\/ 
around the the Banados-Teitelboim-Zanelli black hole (see,
for example, Ref.~\cite{Mann-Solo:1996a}), and the entropy 
of the Boulware-Deser black hole is expected to contain a
power-law correction ($\propto \AHo^{1/3}$) to the
Bekenstein-Hawking entropy (see, for instance, 
Ref.~\cite{Myers-Simo:1988a}). 
It will be worthwhile to investigate as to how the brick wall 
entropy compares with these results. 
We hope to consider these cases in a future publication.

%%%%%%%%%%%%%%%%%%%%%%%%%%%%%%%%%%%%%%%%%%%%%%%%%%%%%%%%%%%%%%%%%%%%%%%%%%%%%%%

\section*{Acknowledgments}

The authors would wish to thank the organizers of The Fifth Meeting on
Field Theoretic Aspects of Gravity that was held at the Birla
Institute of Technology and Science, Goa, India, where this work was
initiated.  We would like to thank Naresh Dadhich for
correspondence and T.~Padmanabhan for discussions.  SSa and LS wish to
thank the Harish-Chandra Research Institute, Allahabad, India, and the
Inter-University Centre for Astronomy and Astrophysics, Pune, India,
respectively, for hospitality, where part of this work was carried
out.  SSa is being supported by the Senior Research Fellowship of the
Council for Scientific and Industrial Research, India.

%%%%%%%%%%%%%%%%%%%%%%%%%%%%%%%%%%%%%%%%%%%%%%%%%%%%%%%%%%%%%%%%%%%%%%%%%%%%%%%

\appendix
\section{Recovering ${\bf \SBH}$ from brick-wall}
\label{app:can-BW}

In the first part of the appendix, we provide the steps leading to the
$\SBH$ in the Schwarzschild-like coordinate system i. e. $f(r) = g(r)$
in the line-element (\ref{eq:spher-tr}). In the second part, we obtain
the same in the tortoise coordinate (\ref{eq:spher-tx}).  As mentioned
in Sec. (\ref{sec:BWall}), the brick-wall model assumes that the the
WKB mode functions are a good approximation for the radial modes near
the horizon. In the case of Schwarzschild-like coordinate system, it
is not apparent whether such an approximation is valid.

%%%%%%%%%%%%%%%%%%%%%%%%%%%%%%%%%%%%%%%%%%%%%%%%%%%%%%%%%%%%%%%%%%%%%%%%%%%%%%%

\subsection{Schwarzschild-like coordinate}

In the case of massless scalar field, the leading order WKB modes 
are given by
\beq
P_0 = \pm \frac{1}{g(r)} \l[E^2 - g(r) \frac{L^2}{r^2} \r]^{1/2} \,  . 
\eeq
Substituting the above expression in Eq. (\ref{eq:Gamma-order}), we get 
\beq
\label{eq:GammaSF}
\Gamma_{0}(E) = \frac{2 E^3}{3 \hbar^3} \, \int\limits_{\rHo+h}^{L} 
\frac{r^2 \, dr}{g^2(r)} \, .
\eeq
Substituting the above expression in (\ref{eq:def-F}) and integrating
over $E$, the free energy $F$ now reads
\beq
\label{eq:FenergySF}
F_{0} =-\frac{2\pi^3}{45 \hbar^3} \frac{1}{\beta^4}\int\limits_{\rHo+h}^{L}
         \frac{r^2}{g^2(r)}dr \, , 
\eeq
and the entropy is 
\beq
\label{eq:EntSF}
S_{0} = \frac{8 \pi^3}{45 \hbar^3} \frac{1}{\beta^3}\int\limits_{\rHo+h}^{L}
         \frac{r^2}{g^2(r)}dr \, . 
\eeq
On expanding the metric near the horizon up to the first-order,
we recover the following standard result~\cite{'tHooft:1984a}
\beq
\label{eq:EntrStd}
S_{0}^{({\rm Std})} = \frac{\rHo^{2} }{90 h_{c}^{2}} \, .
\eeq
However, if we expand the metric to higher orders (\ref{eq:fg-serh}), 
we get
\beq
\label{eq:EntrF}
S_0 = \frac{\rHo^{2} }{90 h_{c}^{2}} +
\l[ \frac{\kappa \rHo}{90} - \frac{g''(\rHo) \rHo^{2}}{360} \r]
\log{\left(\frac{ \rHo^{2}}{ h_{c}^{2}}\right)} \, .
\eeq
It is interesting to note that the form of the brick-wall entropy in
the lowest-order WKB is same as the one obtained in the higher-order
WKB. See Sec. (\ref{sec:BWCorr}) for details.

\subsection{Tortoise coordinates}
\label{app:tortoise}

For the case $f(r) = g(r)$ in (\ref{eq:spher-tr}), the tortoise coordinate
simplifies to 
\begin{eqnarray}
x = \int\limits \frac{dr}{g(r)} \, .
\end{eqnarray}
As mentioned in Sec. (\ref{sec:ST}), in the Rindler approximation
(\ref{eq:rind-met}) the new radial coordinate $x$ takes the form $x =
\log{(r- \rHo)}/g'(\rHo)$. Thus, the horizon $r=\rHo$ corresponds to
$x \rightarrow - \infty$. Hence the tortoise coordinate cover the
range $(-\infty, \infty)$. The massless scalar field propagating in
the background (\ref{eq:spher-tx}) is nothing but the Regge-Wheeler
equation given by
\beq
\frac{d^2 \widehat{R}}{dx^{2}} + \left[\omega^2 - V_{{\rm RW}}[r(x)] \r]
\widehat{R} = 0 
\label{torwave}
\eeq
where 
\beq
\label{eq:rwpot}
V_{{\rm RW}}[r(x)] = \frac{l(l+ 1)}{r^2}~g(r) + \frac{g(r) \, g'(r)}{r} 
\eeq
The Regge-Wheeler potential decays exponentially near the
event-horizon and as a power-law near spatial infinity (for
asymptotically flat spacetimes), i.e.,
\beq
\label{eq:RW-afshor}
V[r(x)] \stackrel{x \to -\infty}{\simeq}  \exp\l(2 \kappa_h x\r)\, ; 
 V[r(x)] \stackrel{x \to \infty}{\simeq} \frac{1}{x^2} \, .
\eeq
Thus, the general solution to Eq. (\ref{torwave}) as $x \to \mp \infty$
can be written as a superposition of plane-waves:
\beq
{\widehat R[x]} \stackrel{x \to \pm\infty}{\sim} 
C_1^{\pm} \exp(i \omega x) + C_2^{\pm} \exp(-i \omega x) \, ,
\label{eq:plawav1}
\eeq
where $C_1^{\pm}, C_2^{\pm}$ are the constants determined by the
choice of the boundary conditions. Thus, in other words, the WKB modes
is a good approximation for the radial modes close to the horizon.

Using the procedure discussed in Sec. (\ref{sec:BWall}), the leading
order density of states is given by
\begin{eqnarray}
\Gamma_0^{(x)}(E) = 
- \frac{2}{3 \pi  \hbar^3} \int\limits_{-H}^{L} dx \, \, \frac{r^2}{g(r)}
\left[E^2-\frac{g(r) g'(r)}{r} \right]^{3/2} \!\!\!\!\! ,  \, 
\end{eqnarray}
where $H$ is a large positive number corresponding to the cutoff
$h$. At the linear order in the near horizon approximation 
%(\ref{eq:fg-rh}), 
we finally
obtain,
\begin{eqnarray}
\Gamma_0^{(x)}(E) &=& -\frac{2}{3 \pi \hbar^3}\int\limits_{-H}^{L}dx \frac{E^3
\rHo^{2}}{g'(\rHo)} \exp{(- 2 \kappa x)} \\ \nonumber 
&=& - \frac{E^3}{6 \pi \hbar^3} \, 
\frac{\rHo^{2}}{\kappa^2} \, \exp{( 2 \kappa H)} \, .
\end{eqnarray}
Following the procedure discussed above, the leading order free energy
and the entropy are given by
\begin{eqnarray}
F_0^{(x)} &=& - \frac{\pi^3 }{90 \hbar^3 \beta^4} \, 
\frac{\rHo^{2}}{\kappa^2} \, \exp{( 2 \kappa H)}\\
S_0^{(x)} &=& \frac{\rHo^2}{180 } \, \kappa \, \exp{(2\kappa H)} \, .
\end{eqnarray}
Now, we want to express this entropy in terms of invariant cutoff
$H_c$ defined as,
\begin{eqnarray}
H_c = \int\limits_{-\infty}^{-H} dr_{*} \sqrt{g(r)} =
\sqrt{\frac{2}{\kappa}}
\exp{(-H\kappa)} \, .
\end{eqnarray}
Using this invariant cutoff, the final expression of entropy is given
by,
\begin{eqnarray}
S_0^{(x)} = \frac{\rHo^{2}}{90 H_{c}^{2}} \, .
\end{eqnarray}
The entropy obtained in the tortoise coordinate is same as obtained in
the Schwarzschild coordinate.

%%%%%%%%%%%%%%%%%%%%%%%%%%%%%%%%%%%%%%%%%%%%%%%%%%%%%%%%%%%%%%%%%%%%%%%%%%%%%%%

\section{Isolating the finite contribution}\label{app:Leibnitz}

In this appendix, we shall outline as to how we isolate the finite 
part of the integrals using the Leibniz rule~
(\ref{lebtz}).

The first integral in the RHS of Eq. (\ref{gamma2}) does not lead to
any divergent term. Using the relation (\ref{lebtz}) --- with $a(\cE)
= 0, b(\cE) = \cE$ --- we get
\br
\label{1stint}
\int\limits_{0}^{\cal E} d\lambda\; P^{(0)}_{2}(r)\; 
\frac{\partial \CGE}{\partial {\cal E}}
&=& \frac{\partial}{\partial {\cal E}}\int\limits_{0}^{\cal E}
d\lambda\; P^{(0)}_{2}(r)\; \CGE \, . \, \, \, \, \, \, \, \, \, \, 
\er
The second and third integral in the RHS of Eq. (\ref{gamma2}) lead to
divergent terms. The origin of the divergent terms can be associated
to the break-down of WKB approximation at the turning point
$(\cE)$. In order to see this, we evaluate the two integrals
explicitly.

Using the relation (\ref{lebtz}) in the second integral of
Eq. (\ref{gamma2}), we get,
\begin{eqnarray}
\label{2ndint}
& & 
\int\limits_{0}^{\cE} \lambda 
\frac{\partial^2 \CGE }{\partial \cE^2} d\lambda  \\
%%%%%%
&=& \frac{\partial}{\partial\cE} 
\int\limits_{0}^{\cE}\lambda \frac{\partial \CGE}{\partial\cE} d\lambda
- \l. \cE \, \frac{\partial \CGE}{\partial \cE} \r|_{\textrm{at}~
\cE = \lambda} \nonumber \\ 
%%%%%
&=&\frac{\partial^2}{\partial \cE^2}\int\limits_{0}^{\cE}\lambda(r) 
\CGE d\lambda
- \l. \frac{\cE}{(\cE - \lambda)^{1/2}} \r|_{\textrm{at}~
\cE = \lambda} \nn \, ,
\end{eqnarray}
where we have used \eq{1stint} to obtain the second equation. 

Using the relation (\ref{lebtz}) in the third integral of
Eq. (\ref{gamma2}), we get
\begin{eqnarray}
\label{3rdint}
\int\limits_{0}^{\cE}\lambda^2 \frac{\partial^3 \CGE}{\partial \cE^3}
d\lambda  &=& \frac{\partial^3}{\partial \cE^3}\int\limits_{0}^{\cE}d\lambda 
\, \, \lambda^2 \, \CGE \\
%%%%
&-& \l. \l[ \frac{\partial}{\partial \cE}\left[\frac{\cE^2}{2 \CGE}\right] -
\frac{\cE^2}{4 \CGEV{3}} \r] \r|_{\cE= \lambda} \, . \nn
\end{eqnarray}
\indent From Eqs. (\ref{2ndint}, \ref{3rdint}), it is clear that both the
integrals have a finite and divergent part. The divergence occurs at
the turning point $\cE = \lambda$. This is not a physical divergences,
this is occurring due to the fact that the WKB approximation is not
valid close to the turning points. However, it can be shown that by
introducing a cutoff close to the turning point the results are
independent of the cutoff.  (For details, see Sec. (10.7) in
Ref. \cite{Bender-Orza:1978}.)

%%%%%%%%%%%%%%%%%%%%%%%%%%%%%%%%%%%%%%%%%%%%%%%%%%%%%%%%%%%%%%%%%%%%%%%%%%%%%%%

\begin{widetext}

\section{Second order contribution when ${\bf f(r) \neq g(r)}$}
\label{app:corr-genmetric}

Earlier, in Subsection~\ref{sec:4D-IIorder}, we had computed the
brick wall entropy at the second order in the WKB approximation 
for the specific case wherein $f(r)=g(r)$ in the 
line-element~(\ref{eq:spher-tr}).
In this appendix, we shall obtain the corresponding result for
the more general case of $f(r)\neq g(r)$.
For simplicity, we shall again consider a massless field. 

When $f(r)\neq g(r)$, we find that, the `momentum' at the second
order~$P_2(r)$ [cf.~Eq.~(\ref{eq:for-P2})] can be written as
\beq
P_2^{(G)}(r) 
= \l(\frac{P^{(0)}_{2G} (r)}{\CGE}\r)
+ \lambda(r)\, \l(\frac{ P^{(1)}_{2G}(r)}{\CGEV{3}}\r)
+\; \lambda^{2}(r)\, \l(\frac{P^{(2)}_{2G}(r)}{\CGEV{5}}\r)\,,
\label{eq:p2-fn=g}
\eeq
where ${\cal G}({\cal E},r)$ is given by Eq.~(\ref{eq:def-CGE}). 
We have defined the functions $P^{(0)}_{2G}(r)$, $P^{(1)}_{2G}(r)$
and $P^{(2)}_{2G}(r)$ to be
\br
P^{(0)}_{2G}(r) 
&=& -\l(\frac{G(r)\, f'(r)}{4\, r\, f(r)}\r)
-\l(\frac{G(r)\, g'(r)}{4\, r\, g(r)}\r)\,, \nn\\ 
P^{(1)}_{2G}(r)
&=&\l(\frac{3\, G(r)\, f(r)}{4 r^4}\r)
-\l(\frac{5\, G(r)\, f'(r)}{8\, r^3}\r) 
+\l(\frac{G(r)\, f'(r)^2}{16\, r^2\, f(r)}\r)
-\l(\frac{G(r)\, f(r)\, g'(r)}{8 r^3 g(r)}\r)
+ \l(\frac{G(r)\, f'(r)\, g'(r)}{16\, r^2\, g(r)}\r)
+\l(\frac{G(r)\, f''(r)}{8\, r^2}\r)\,,\nn\\
P^{(2)}_{2G}(r) 
&=&  \l(\frac{5\, G(r)\,f(r)^2}{8\, r^6}\r)
-\l(\frac{5\, G(r)\, f(r)\, f'(r)}{8\, r^5}\r)
+\l(\frac{5\, G(r)\, f'(r)^2}{32\, r^4}\r) \, ,
\er
with the quantity $G(r)$ given by Eq.~(\ref{eq:defG}).
On  we find that the number of 
states at the second order~$\Gamma_{2}^{(G)}(E)$ can be expressed
as follows:
\beq
\Gamma_{2}^{(G)} (E) 
= \l(\frac{1}{\pi\, \hbar}\r)\,   
\int\limits_{\rHo +h}^{\cal L}\! dr\;
\left(2\, P^{(0)}_{2G}(r)\; \frac{\partial}{\partial \cE} 
\int\limits_{0}^{\cE}\! d\lambda\; \CGE 
- 4\, P^{(1)}_{2G}(r)\; \frac{\partial^2}{\partial \cE^2}
\int\limits_{0}^{\cE}\! d\lambda\; \lambda\; \CGE 
+ \l(\frac{8}{3}\r)\, P^{(2)}_{2G}(r)\; \frac{\partial^3}{\partial
\cE^3}\int\limits_{0}^{\cE}\! d\lambda\; \lambda^2\; \CGE\right) \, , 
\label{eq:Gamma-fneqg}
\eeq
Repeating the procedure discussed in the last appendix to identify 
and ignore the divergences in the above expression and substituting this expression in the integral~(\ref{eq:def-F}) for the 
free energy, we obtain that 
\br
F_{2}^{(G)} 
&=& \l(\frac{\pi}{36\, \beta\, \hbar}\r)\,
\int\limits_{_{\rHo +h}}^{\cal L}\! 
\frac{dr\, G(r)}{g(r)\, f^{3}(r)}\,
\left[4\, f^{2}(r)\, g(r)
- 16\, r\, f(r)\, f'(r)\, g(r)
+ 7\, r^2\, {f'}^{2}(r)\, g(r)\r.\nn\\ 
& &\qquad\qquad\qquad\qquad\qquad
\qquad\qquad\qquad\qquad\qquad\qquad\qquad\qquad
\l.-\, 3\, r^2\, f(r)\, f'(r)\, g'(r)
- 6\, r^2\, f(r)\, g(r)\, f''(r)\right].\qquad\qquad
\er
Using the relation~(\ref{eq:CE-entropy}) and expanding the functions
$f(r)$ an $g(r)$ about the event horizon as in Eq.~(\ref{eq:fg-serh}), 
we obtain the the expression for the entropy to be
\beq
S_{2}^{(G)}= \l(\frac{\rHo^2}{9\, h_{c}^{2}}\r)
- \left[\l(\frac{g''(\rHo)}{144}\r)
+\l(\frac{f''(\rHo)\, g'(\rHo)}{48\, f'(\rHo)}\r)
+ \l(\frac{4\, \pi\,\rHo}{9\, \beta\, \hbar}\r)\, 
\sqrt{\frac{g'(\rHo)}{f'(\rHo)}}
\right] \log\l(\frac{\rHo^2}{h_c^2}\r).
\eeq
%

%%%%%%%%%%%%%%%%%%%%%%%%%%%%%%%%%%%%%%%%%%%%%%%%%%%%%%%%%%%%%%%%%%%%%%%%%%%%%%%

\section{Second order contribution for a massive 
field}\label{app:corr-mass}

In Subsection~\ref{sec:4D-IIorder}, we had evaluated the brick wall
entropy at the second order for a massless field in four spacetime 
dimensions. 
We had considered the specific case wherein $f(r) = g(r)$ in the 
line-element~(\ref{eq:spher-tr}).
In this appendix, we shall discuss the corresponding result for a 
massive field.

For a massive field, the quantity~$P_{2}(r)$ as given by 
Eq.~(\ref{eq:for-P2}) can be expressed as follows:
\beq
\label{eq:p2-f=gm}
P_2^{(m)} (r) 
= \l(\frac{P^{(0)}_2 (r)}{\CGmE}\r)
+ \l(\frac{\lambda(r)\, P^{(1)}_{2}(r) + P^{(0)}_{2m}(r)}{\CGmEV{3}}\r)
+ \l(\frac{\lambda^{2}(r)\, P^{(2)}_{2}(r) 
+ \lambda(r)\, \l[P^{(1)}_{2m}(r)+P^{(2)}_{2m}(r)\r]}{\CGmEV{5}}\r)\,, 
\eeq
where we have defined the functions $P^{(0)}_{2m}(r)$,  $P^{(1)}_{2m}(r)$ 
and $P^{(2)}_{2m}(r)$ to be
\beq
P^{(0)}_{2m}(r) =\l(\frac{m^2}{8}\r)
\l[{g'}^{2}(r) + g(r)\, g''(r)\r],\;\;
P^{(1)}_{2m}(r) = \l(\frac{5\, m^2\, g(r)\, g'(r)}{16\, r^3}\r)
\l[2\, g(r)+r\, g'(r)\r]\;\;{\rm and}\;\;
%%%
P^{(2)}_{2m}(r)=  \l(\frac{5}{32}\r) 
\l[m^4\, g(r)\, {g'}^{2}(r)\r]\,,
\eeq
with the quantity ${\cal G}_{m}({\cal E},r)$ being given by
\beq
{\cal G}_{m}({\cal E},r) = \l[{\cal E} - \lambda(r)- m^2 g(r) \r]^{1/2}.
\eeq
Carrying out the procedure we had discussed in Appendix~\ref{app:Leibnitz} 
to identify and discard the divergences, we obtain the second order 
density of states $\Gamma^{(m)}_2(E)$ for the massive field to be
\begin{eqnarray}
\Gamma_{2}^{(m)} (E) 
&=& \l(\frac{1}{2\, \hbar\, \pi}\r)\,
\int\limits_{_{\rHo + h}}^{L}\! dr\, r^2\, 
\l[P^{(0)}_{2}(r)\; \frac{\partial}{\partial \cE} 
\int\limits_{0}^{\cE_{m}}\! d\lambda\; \CGmE 
- 2\, P^{(1)}_{2}(r)\; \frac{\partial^2}{\partial \cE^2}
\int\limits_{0}^{\cE_m}\! d\lambda\; \lambda \; \CGmE \r.\nn\\ 
& &\qquad\qquad\qquad\qquad\qquad\qquad\qquad
+\l. 4\, P^{(0)}_{2m}(r)\; \frac{\partial^2}{\partial \cE^2}
\int\limits_{0}^{\cE_{m}}\! d\lambda\; \CGmE 
+  \l(\frac{4}{3}\r)\, P^{(2)}_{2}(r)\;
\frac{\partial^3}{\partial \cE^3} 
\int\limits_{0}^{\cE_{m}}\! d\lambda\; \lambda^2\;\CGmE\r.\nn\\
& &\qquad\qquad\qquad\qquad\qquad\qquad\qquad
+\l.\l(\frac{8}{3}\r)\, \l[P^{(1)}_{2m}(r)+P^{(2)}_{2m}(r)\r]\,
\frac{\partial^3}{\partial \cE^3}
\int\limits_{0}^{\cE_{m}}\! d\lambda\; \lambda\; \CGmE\r]\,,
\end{eqnarray}
where $\cE_{m} = \l[{\cal E}-m^2\, g(r)\r]$.
As in the massless case, on expanding the function $g(r)$ near 
the event horizon as in Eq. (\ref{eq:fg-serh}), we find that 
$\Gamma^{(m)}_2(E)$ can be expressed as
\br
\Gamma_{2}^{(m)}(E) 
&=& \l(\frac{1}{\pi\, \hbar \pi}\r)\;
\int\limits_{\rHo+h}^{L}\!dr\,
\l[-\l(\frac{2\, E\, \rHo}{3\, (r-\rHo)}\r)
+\l(\frac{E\, \rHo^2}{3\, (r-\rHo)^2}\r)
-\l(\frac{E\, \rHo^2\, g''(\rHo)}{12\, (r-\rHo)\, \kappa}\r)
+\l(\frac{m^2\, \rHo^2\, \kappa}{E\, (r-\rHo)}\r)\right].
\er
Note that the last term containing the mass~$m$ of the field is 
inversely proportional to~$E$.
Recall that, in Subsection~\ref{sec:4D-IVorder}, we had encountered
such a behavior at the fourth order for the massless field
[cf. Eq.~(\ref{eq:Gamma4Fin})]. 
As we had then pointed out, such a dependence on the energy~$E$ in
the number of states leads to a free energy that turns out to be 
independent of the inverse temperature~$\beta$ and, hence, the term 
does not contribute to the entropy.
Therefore, the massless and the massive fields lead to the same 
entropy at the second order in the WKB approximation.

%%%%%%%%%%%%%%%%%%%%%%%%%%%%%%%%%%%%%%%%%%%%%%%%%%%%%%%%%%%%%%%%%%%%%%%%%%%%%%%

\section{Explicit forms of ${\bf \PiIV}$}\label{app:PiIV}

The functions $P^{(i)}_4(r)$ [where $i=0,\cdots,4$] are given by,
\begin{eqnarray}
P^{(0)}_4(r)&=& -\frac{5}{2} g(r) P^{(0)}_2 (r)^2 
- \frac{g(r)}{r} \, g'(r) \, P^{(0)}_2 (r) 
- \frac{1}{4} g'(r)^2 P^{(0)}_2 (r)  \nonumber \\ 
& & - \frac{3}{4} g(r) g'(r) P'^{(0)}_2 (r)-\frac{1}{4} g(r) P^{(0)}_2 (r)
g''(r)-\frac{1}{4} g(r)^2 P''^{(0)}_2 (r) \, , 
\er
%%%%
\br
P^{(1)}_4 (r) &=& - 5g(r) P^{(0)}_2 (r) P^{(1)}_2( r)
+ \frac{5}{4 r^3} \l[g(r)^2 \, g'(r) \, P^{(0)}_2 (r) + g(r)^3 P'^{(0)}_2 (r)\r]
\nonumber \\
&& -\frac{5}{8 r^2}\l[g(r) P^{(0)}_2 (r) g'(r)^2 + g(r)^2 g'(r)
P'^{(0)}_2 (r)\r] 
-\frac{1}{r} g(r) g'(r) P^{(1)}_2 (r) 
\nonumber \\ 
&& - \frac{1}{4} 
\l[ g'(r)^2 P^{(1)}_2 (r) + 3 g(r) g'(r) P'^{(1)}_2 (r) 
+ g(r) g''(r) P^{(1)}_2(r) + g(r)^2 P''^{(1)}_2 (r) \r] \, , 
\er
%%%%%
\br
P^{(2)}_4 (r) &=& 
- 5 g(r) P^{(0)}_2(r) P^{(2)}_2(r) 
- \frac{5}{2} g(r) P^{(1)}_2(r)^2 
- \frac{3}{16} g(r) g'(r) P'^{(2)}_2(r) \nn \\
& & - \frac{1}{4} \l[g(r) g''(r) P^{(2)}_2 (r) 
+ g(r)^2 P''^{(2)}_2 (r) 
+ g'(r)^2 P^{(2)}_2 (r) \r]
-\frac{5}{4 r^6} g(r)^4 P^{(0)}_2(r)  \nonumber\\ 
&&
+ \frac{5}{4 r^5} g(r)^3 g'(r) P^{(0)}_2(r) 
-\frac{1}{2 r^4} \l[3 g(r)^3 P^{(1)}_2(r)
- \frac{5}{8} g(r)^2 g'(r)^2 P^{(0)}_2 (r)\r] \nn \\
& & 
+\frac{1}{4 r^3} \l[15 g(r)^2 g'(r) P^{(1)}_2 (r) 
+ 9 g(r)^3 P'^{(1)}_2(r) \r] 
-\frac{1}{r} g(r) g'(r) P^{(2)}_2(r) \nn \\
& & 
- \frac{1}{8 r^2} \l[11 g'(r)^2 g(r) P^{(1)}_2 (r) 
+ 9 g(r)^2 g'(r) P'^{(1)}_2 (r) 
+ 2 g(r)^2 g''(r) P^{(1)}_2 (r) \r]
\er
%%%%%%
\br
P^{(3)}_4 (r) &=&
- 5 g(r) P^{(1)}_2 (r) P^{(2)}_2(r) 
-\frac{23}{4 r^6} g(r)^4 P^{(1)}_2 (r)
+ \frac{23 g(r)^3 P^{(1)}_2(r) g'(r)}{4 r^5}
\nonumber \\ 
&& -\frac{1}{r^4} 
\l[ 3 g(r)^3 P^{(2)}_2 (r) - \frac{23}{16} g(r)^2 g'(r)^2 P^{(1)}_2(r) \r]
\nonumber \\ 
&& + \frac{1}{4 r^3}  \l[25 g(r)^2 P^{(2)}_2(r) g'(r)
+ 13 g(r)^3 P'^{(2)}_2(r) \r]  
\nonumber \\
& & 
- \frac{1}{8 r^2} \l[17 g(r) P^{(2)}_2(r) g'(r)^2 
+ 13 g(r)^2 g'(r) P'^{(2)}_2(r) 
+ 4 g(r)^2 P^{(2)}_2(r) g''(r) \r] \, , 
\er
%%%%%
\br
P^{(4)}_4 (r) &=& -\frac{49}{4 r^6} g(r)^4 P^{(2)}_2(r)
+\frac{49}{4 r^5} g(r)^3 g'(r) P^{(2)}_2(r) 
-\frac{5}{2} g(r) P^{(2)}_2 (r)^2 
- \frac{49 }{16 r^4} g(r)^2  g'(r)^2 P^{(2)}_2(r) \, ,
\end{eqnarray}
\br
\label{eq:SigIVorder}
\Sigma^{(4)}(r) &=& 
\frac{323}{10080} \, \frac{r^2 g'(r)^4}{g(r)^2} 
+ \frac{101 r^2 g'(r)^2 g''(r)- 631 \, r \, g'(r)^3}{1680 \, g(r)} \nn \\
%%%%
& & 
+ \frac{7 r^2}{840} \l[g''(r)^2  + 7 g'(r) g^{(3)}(r) + 5 g^{(4)}(r) g(r) \r]
+ \frac{r}{840} \l[ 155 g'(r) g''(r)  + 252 g^{(3)}(r) g(r) \r] \nn \\
%%%%
& & 
+ \frac{467 g'(r)^2 + 150 g''(r) g(r)}{420} 
+ \frac{17}{630} \, \frac{g(r)^2}{r^2} 
- \frac{1223}{2520} \frac{g'(r) g(r)}{r} 
\er
%
%%%%%%%%%%%%%%%%%%%%%%%%%%%%%%%%%%%%%%%%%%%%%%%%%%%%%%
%%%%%%%%%%%%% NEW APPENDIX %%%%%%%%%%%%%%%%%%%%%%%%%%%
%%%%%%%%%%%%%%%%%%%%%%%%%%%%%%%%%%%%%%%%%%%%%%%%%%%%%%
\section{Explicit forms of ${\bf \PiVI}$}
\label{app:PiVI}

The functions $P^{(i)}_6(r)$ [where $i=0,\cdots,6$] are given by,
{\small
\begin{eqnarray}
P^{(0)}_6 (r) &=& -2 g(r)^2 P^{(0)}_2(r)^3 
- 5 g(r) P^{(0)}_2(r)P^{(0)}_4(r) 
- \frac{g(r) g'(r)}{2 r} \l[ g(r) P^{(0)}_2(r)^2 + P^{(0)}_4(r) \r] 
 \nonumber  \\
& & 
- \frac{g(r)}{8} \l[2 g(r)^2 P^{(0)}_2(r) P''^{(0)}_2(r) 
-  g'(r)^2  P^{(0)}_2(r)^2  
- 3 g(r)^2 P'^{(0)}_2(r)^2 
+ 6  g'(r) P'^{(0)}_4(r) \r. \nn \\
& & \l. \qquad \quad + 2 g(r) g''(r) P^{(0)}_2(r)^2 
+ 2 g(r) P''^{(0)}_4(r) + 2 g''(r)  P^{(0)}_4(r) \r]  
- \frac{1}{4} P^{(0)}_4(r) g'(r)^2 
\er
%%%%%%%%
\br
P^{(1)}_6 (r) &=& 
-\frac{3}{4 r^4} g(r)^3 \l[ g(r) P^{(0)}_2(r)^2 + 2 P^{(0)}_4(r) \r] 
- \frac{g(r)}{r} \l[g(r) g'(r) P^{(0)}_2(r) P^{(1)}_2(r) 
+ g'(r) P^{(1)}_4(r) \r] \nn \\
& & 
- \frac{g(r)^2 }{4 r^3} \l[2 g(r) P^{(0)}_2(r)^2 g'(r)
- 15 g'(r)P^{(0)}_4(r) 
+ g(r)^2 P^{(0)}_2(r) P'^{(0)}_2(r)
- 9 g(r) P'^{(0)}_4(r) \r] \nn \\
& & 
+ \frac{g(r)}{8 r^2} \l[g(r)^2 g'(r) P^{(0)}_2(r)  P'^{(0)}_2(r) 
- 11 g'(r)^2 P^{(0)}_4(r) 
- 9  g(r) g'(r) P'^{(0)}_4(r)
-  g(r)^2 g''(r) P^{(0)}_2(r)^2 \r] \nn \\
%%%
& & 
- \frac{1}{4 r^2} g(r)^3 g''(r) P^{(0)}_4(r)  
- 6 g(r)^2 P^{(0)}_2(r)^2 P^{(1)}_2(r) 
- 5 g(r) P^{(1)}_2(r) P^{(0)}_4(r) 
- 5 g(r) P^{(0)}_2(r) P^{(1)}_4(r) \nn \\
%%%
& &
+ \frac{g(r) }{4} \l[g'(r)^2  P^{(0)}_2(r) P^{(1)}_2(r) 
+ 3 g(r)^2 P'^{(0)}_2(r) P'^{(1)}_2(r) 
- g(r)^2  P^{(1)}_2(r) P''^{(0)}_2(r) 
- g(r)^2 P^{(0)}_2(r) P''^{(1)}_2(r)  \r. \nn \\
%%%%
& & 
\l. - 2 g(r) g''(r) P^{(0)}_2(r) P^{(1)}_2(r) \r]
- \frac{1}{4} \l[g'(r)^2 P^{(1)}_4(r)  
+ g(r)^2 P''^{(1)}_4(r) 
+  g(r) P^{(1)}_4(r) g''(r) 
+ 3 g(r) g'(r) P'^{(1)}_4(r) \r]
\er
%%%%%%%%%%%%%%%%%%
{\small
\br
P^{(2)}_6(r) &=& 
-\frac{g(r)^4}{8 r^6} \l[3 g(r) P^{(0)}_2(r)^2 + 23 P^{(0)}_4(r) \r]
+ \frac{g(r)^3 g'(r)}{8 r^5} \l[3 g(r) P^{(0)}_2(r)^2 + 46  P^{(0)}_4(r) \r] 
\nn \\
& & 
- \frac{3 g(r)^3 }{r^4} \l[g(r) P^{(0)}_2(r) P^{(1)}_2(r) + P^{(1)}_4(r) \r]
- \frac{g(r)^2 g'(r)^2}{32 r^4}  \l[3 g(r)P^{(0)}_2(r)^2 + 46  P^{(0)}_4(r) \r]
\nn \\
& &  
- \frac{g(r) g''(r) }{4} 
\l[ P^{(2)}_4(r) + 2 g(r) P^{(0)}_2(r)P^{(2)}_2(r) +
g(r) g''(r) P^{(1)}_2(r)^2 \r]
+ \frac{g(r)^2 }{4 r^3} \l[ 3 P^{(0)}_2(r) P'^{(1)}_2(r) 
+ 13 g(r) P'^{(1)}_4(r) \r] 
\nn \\
& & + \frac{g(r)^2 }{4 r^3} \l[8 g(r) g'(r) P^{(0)}_2(r) P^{(1)}_2(r) 
+ 25   g'(r) P^{(1)}_4(r) 
- 7  g(r)^2 P^{(1)}_2(r) P'^{(0)}_2(r) \r]
\nn \\
& & + \frac{g(r) g'(r)}{8 r^2} 
\l[ - 17 g'(r) P^{(1)}_4(r) 
+ 7  g(r)^2  P^{(1)}_2(r) P'^{(0)}_2(r)
- 3 g(r)^2  P^{(0)}_2(r)  P'^{(1)}_2(r)
- 13  g(r) P'^{(1)}_4(r) \r] \nn \\
& & 
- \frac{g(r)^2 g''(r) }{2 r^2}  \l[ P^{(0)}_2(r) P^{(1)}_2(r) +
P^{(1)}_4(r) \r] 
 - \frac{g(r) g'(r)}{2 r} \l[ g(r) P^{(1)}_2(r)^2 
+ 2 g(r) P^{(0)}_2(r) P^{(2)}_2(r) 
+  P^{(2)}_4(r) \r] \nn \\
& & - 6 g(r)^2 \l[P^{(0)}_2(r) P^{(1)}_2(r)^2 + P^{(0)}_2(r)^2 P^{(2)}_2(r) \r]
- 5 g(r) \l[P^{(2)}_2(r) P^{(0)}_4(r) +  P^{(1)}_2(r) P^{(1)}_4(r)
+ P^{(0)}_2(r) P^{(2)}_4(r) \r] \nn \\
& & 
+ \frac{g(r) g'(r)^2 }{8} 
\l[P^{(1)}_2(r)^2 
+ P^{(0)}_2(r) P^{(2)}_2(r) 
- \frac{6}{g(r)} P'^{(2)}_4(r) 
- \frac{2}{g(r)} P^{(2)}_4(r) \r] \\
& & 
-\frac{1}{4} g(r)^3 
\l[ P^{(0)}_2(r) P''^{(2)}_2(r)
+ P^{(1)}_2(r) P''^{(1)}_2(r) 
+ \frac{P''^{(2)}_4(r)}{g(r)} 
+ P^{(2)}_2(r) P''^{(0)}_2(r) 
+ \frac{3}{2} P'^{(1)}_2(r)^2 
+ 3 P'^{(0)}_2(r) P'^{(2)}_2(r) \r] \nn
\er
}
%%%%%%%%%%%%%%%
{\small
\begin{eqnarray}
P^{(3)}_6(r) &=& 
-\frac{g(r)^4 }{4 r^6} \l[9 g(r) P^{(0)}_2(r) P^{(1)}_2(r) + 49 P^{(1)}_4(r)
\r]
+ \frac{g'(r)g(r)^3}{4 r^5}  \l[9 g(r) P^{(0)}_2(r) P^{(1)}_2(r) + 49
P^{(1)}_4(r) \r] \nn \\
& & 
-\frac{9}{4 r^4} g(r)^4 \l[P^{(1)}_2(r)^2
+ 2 P^{(0)}_2(r) P^{(2)}_2(r) + \frac{2}{g(r)} P^{(2)}_4(r) \r] 
- \frac{g(r)^2 g'(r)^2 }{16 r^4} \l[9 g(r) P^{(0)}_2(r) P^{(1)}_2(r) 
+ 49  P^{(1)}_4(r) \r] \nonumber \\
& &
- \frac{g(r)^4}{4 r^3}  \l[13 P^{(2)}_2(r) P'^{(0)}_2(r)
- 7  P^{(0)}_2(r) P'^{(2)}_2(r)
+ 3 g(r)^4 P^{(1)}_2(r) P'^{(1)}_2(r) 
- \frac{17}{g(r)} P'^{(2)}_4(r) \r] \nn \\
%%%
& & 
+ \frac{g(r)^3 g'(r)}{4 r^3} \l[\frac{35}{g(r)} P^{(2)}_4(r) 
+ 6 P^{(1)}_2(r)^2 + 12  P^{(0)}_2(r)P^{(2)}_2(r) \r]
- \frac{g(r)}{4}  \l[3 g'(r) P'^{(3)}_4(r) + g''(r)  P^{(3)}_4(r) \r] \nn \\
%%%
& & 
- \frac{3}{8 r^2}  g(r)^3 g''(r) 
\l[P^{(1)}_2(r)^2 + 2 P^{(0)}_2(r) P^{(2)}_2(r) 
- \frac{2}{g(r)} P^{(2)}_4(r) \r] 
- \frac{g(r) g'(r)^2}{8 r^2}  \l[ 23 g'(r) P^{(2)}_4(r) + 17 g(r)
P'^{(2)}_4(r)\r]
\nonumber \\
& & 
+ \frac{g(r)^3 g'(r)}{8 r^2}  \l[13 P^{(2)}_2(r)  P'^{(0)}_2(r) 
+ 3 P^{(1)}_2(r)  P'^{(1)}_2(r) - 7  P^{(0)}_2(r)  P'^{(2)}_2(r) \r]
\nonumber \\ 
& & 
- \frac{g(r) g'(r)}{r}  \l[g(r) P^{(1)}_2(r) P^{(2)}_2(r) + P^{(3)}_4(r) \r]
- 2 g(r)^2 P^{(1)}_2(r)^3 
- 12 g(r)^2 P^{(0)}_2(r) P^{(1)}_2(r) P^{(2)}_2(r)
- \frac{1}{4}g(r)^2 P''^{(3)}_4(r) 
\nn \\
& & 
- 5 g(r) \l[P^{(2)}_2(r) P^{(1)}_4(r) + P^{(1)}_2(r) P^{(2)}_4(r) 
+ P^{(0)}_2(r) P^{(3)}_4(r) \r]
+ \frac{g(r) g'(r)^2 }{4} \l[g(r) P^{(1)}_2(r) P^{(2)}_2(r) 
- P^{(3)}_4(r) \r]
\nn \\
& & 
+ \frac{g(r)^3 }{4} \l[ 3 P'^{(1)}_2(r) P'^{(2)}_2(r)
- 2 P^{(1)}_2(r) P''^{(2)}_2(r)
- 2  P^{(2)}_2(r) P''^{(1)}_2(r)
- P^{(1)}_2(r) P''^{(2)}_2(r) \r]
\er
%%%%%%%%%%%%%%%%%%%%%%%%%
\br 
P^{(4)}_6 (r) &=& 
-\frac{g(r)^5 }{8 r^6}  \l[3  P^{(1)}_2(r)^2
+ 46 P^{(0)}_2(r)P^{(2)}_2(r)
+ \frac{166}{g(r)} P^{(2)}_4(r)\r] 
- \frac{g(r)^2 g''(r)}{r^2} \l[g(r) P^{(1)}_2(r)P^{(2)}_2(r) 
+  P^{(3)}_4(r)  \r] 
\nn \\
& & 
+ \frac{g'(r) g(r)^4}{8 r^5}  \l[46 P^{(0)}_2(r)P^{(2)}_2(r) 
+ 3  P^{(1)}_2(r)^2 
+ \frac{166}{g(r)} P^{(2)}_4(r) \r]
- \frac{6 g(r)^3}{r^4} \l[g(r) P^{(2)}_{1}(r)P^{(2)}_2(r) + P^{(3)}_4(r)
\r] \nonumber \\ 
%%%
& & 
- \frac{g(r)^3 g'(r)^2 }{32 r^4}
\l[3 P^{(1)}_2(r)^2 + 46 P^{(0)}_2(r)P^{(2)}_2(r) 
+ \frac{166}{g(r)} P^{(2)}_4(r) \r]
- \frac{g'(r) g(r)}{2 r} \l[g(r) P^{(2)}_2(r)^2 + 2 P^{(4)}_4(r)\r]
\nonumber \\ 
& & 
+ \frac{g(r)^2}{4 r^3}  \l[16 g(r) g'(r)P^{(1)}_2(r) P^{(2)}_{2}(r) 
+ 45 g(r) P^{(3)}_4(r) + 84 g(r) P'^{(3)}_4(r) \r] \nn \\
%%%
& & 
- \frac{g(r)^4}{4 r^3}  \l[9 P^{(2)}_2(r)P'^{(1)}_2(r)
+ g(r)^4 P^{(1)}_2(r)P'^{(2)}_2(r)\r] 
-\frac{g(r) g''(r) }{4} \l[g(r) P^{(2)}_2(r)^2 + P^{(4)}_4(r) \r]
\nn \\
& & 
+ \frac{g(r)^3 g'(r)}{8 r^2}  \l[ 9 P^{(2)}_2(r) P'^{(1)}_2(r)
-  P^{(1)}_2(r) P'^{(2)}_2(r)  
- \frac{21}{g(r)} P'^{(3)}_4(r)
- 29 \frac{g'(r)}{g(r)} P^{(3)}_4(r) \r] \nn \\
%%%%
&& 
- 6g(r)^2 \l[P^{(1)}_2(r)^2P^{(2)}_2(r) + P^{(0)}_2(r)P^{(2)}_2(r)^2 \r]
- 5 g(r) \l[ P^{(2)}_2(r) P^{(2)}_4(r) + P^{(1)}_2(r)P^{(3)}_4(r)
+ P^{(0)}_2(r) P^{(4)}_4(r) \r]  \\ 
&& 
+ \frac{1}{8} g(r)g'(r) \l[g'(r) P^{(2)}_2(r)^2 
- 2 \frac{g'(r)}{g(r)} P^{(4)}_4(r) 
- 6 P'^{(4)}_4(r) \r]
+ \frac{1}{8} g(r)^3 \l[ P'^{(2)}_2(r)^2
- 2  P^{(2)}_2(r)P''^{(2)}_2(r) 
- \frac{2}{g(r)} P''^{(4)}_4(r) \r] \nn 
\er
%
%%%%%%%%%%%
\br
P^{(5)}_6(r) &=& 
-\frac{5 g(r)^4}{4 r^6}   \l[g(r) P^{(1)}_2(r) P^{(2)}_2(r)
+ 25  P^{(3)}_4(r) \r]
+ \frac{5 g'(r) g(r)^2 }{4 r^5}  \l[g(r) P^{(1)}_2(r) P^{(2)}_2(r) 
+ 25 P^{(3)}_4(r) \r]
\nonumber \\
& & 
-\frac{15}{4 r^4} g(r)^3 \l[ g(r) P^{(2)}_2(r)^2 + 2 P^{(4)}_4(r) \r]
- \frac{5}{16 r^4} g(r)^2 g'(r)^2 \l[g(r) P^{(1)}_2(r) P^{(2)}_2(r) 
+ 25 P^{(3)}_4(r) \r]
\nonumber \\ 
& &
+ \frac{5 g(r)^2 g'(r)}{4 r^3}  \l[2 g(r)P^{(2)}_2(r)^2 
+ 11 g(r)^2  P^{(4)}_4(r) \r] 
-\frac{5 g(r)^3}{4 r^3}  \l[g(r) P^{(2)}_2(r) P'^{(2)}_2(r)
+ 25 P'^{(4)}_4(r) \r] \nonumber \\
& & 
- \frac{35 g(r) g'(r)^2}{8 r^2}  P^{(4)}_4(r) 
+ \frac{5 g(r)^3 g'(r) }{8 r^2}  \l[g(r) P^{(2)}_2(r) P'^{(2)}_2(r) 
- 5 P'^{(4)}_4(r) \r] 
- 6 g(r)^2 P^{(1)}_2(r) P^{(2)}_2(r)^2
\nn \\
& & 
- \frac{5 g''(r) g(r)^2 }{8 r^2} \l[g(r) P^{(2)}_2(r)^2 
- 2P^{(4)}_4(r) \r] 
- 5 g(r) \l[ P^{(2)}_2(r) P^{(3)}_4(r)
 + P^{(1)}_2(r) P^{(4)}_4(r) \r]
\er
%%%%%%%%%%%
\br
P^{(6)}_6 (r) &=& 
\frac{5}{8 r^6}  g(r)^4 \l[ g(r)P^{(2)}_2(r)^2 - 70 P^{(4)}_4(r) \r]
- \frac{5 g(r)^3 g'(r)}{8 r^5} \l[g(r) P^{(2)}_2(r)^2 
- 70  P^{(4)}_4(r) \r] \nonumber \\
& & 
+ \frac{5 g(r)^2 g'(r)^2}{32 r^4} \l[ g(r) P^{(2)}_2(r)^2
- 70  P^{(4)}_4(r)  \r]
- 2 g(r)^2 P^{(2)}_2(r)^3 
- 5 g(r) P^{(2)}_2(r) P^{(4)}_4(r)
\er
}
\br
\label{eq:SigVIorder}
\Sigma^{(6)}(r) &=& 
\frac{9341 g(r)^4}{180180 \, r^4}
-\frac{4741 g'(r) g(r)^3}{16380 \, r^3}
+ \frac{1308784 \, g''(r) g(r)^3+3926504 \, g'(r)^2 g(r)^2}{5765760 \,
r^2} \\
%%%%%%
& & 
- \frac{536120 g^{(3)}(r) \, g(r)^3 +3719032 \, g'(r) \, g''(r) g(r)^3 
 + 2869040 \, g'(r)^3 g(r)}{5765760 \, r} \nn \\
%%%%%%
& & 
+ \frac{213928 g^{(4)}(r) g(r)^3+761280 g''(r)^2 g(r)^2 
   + 1261208 g'(r) g^{(3)}(r) g(r)^2 +1508748 g'(r)^2 g''(r) g(r) 
   + 435674 g'(r)^4}{5765760} \nn \\
%%%%%%
& & 
+\frac{r \left(137280 g^{(5)}(r) g(r)^3 
           + 947804 g''(r) g^{(3)}(r) g(r)^2 
           + 903188 g'(r) g^{(4)}(r) g(r)^2 
           + 462228 g'(r) g''(r)^2 g(r) \r)}{5765760}\nn \\
%%%%%%
& & 
+\frac{r\l(971568 g'(r)^2 g^{(3)}(r) g(r)^2 
           - 4496 g'(r)^3 g''(r) g(r) 
           - 24496 g'(r)^5 \right)}{5765760 \, g(r)} \nn \\
%%%%%%
& & +\frac{r^2 \left(3473 g'(r)^6
                   -  27895 g(r) g''(r) g'(r)^4 
                   + 84032 g(r)^2 g^{(3)}(r) g'(r)^3
                   + 113100 g(r)^2 g''(r)^2 g'(r)^2 \r)}{5765760 g(r)^2} \nn \\
%%%%%%
& & +\frac{r^2 \left(183898 g(r)^3 g^{(4)}(r) g'(r)^2 
                   + 316316 g(r)^3 g''(r) g^{(3)}(r) g'(r)
                   + 99528 g(r)^4 g^{(5)}(r) g'(r) \r)}{5765760 g(r)^2} \nn \\
%%%%%%
& & +\frac{r^2 \left(40040 g(r)^3 g''(r)^3 
                   + 61776 g(r)^4 g^{(3)}(r)^2 
                   + 123552 g(r)^4 g''(r) g^{(4)}(r) 
                   + 12012 g(r)^5 g^{(6)}(r)\right)}{5765760 g(r)^2} \nn 
\er
}
\end{widetext}

%%%%%%%%%%%%%%%%%%%%%%%%%%%%%%%%%%%%%%%%%%%%%%%%%%%%%%%%%%%%%%%%%%%%%%%%%%%%%%%

%%%%%%%%%%%%%%%%%%%%%%%%%%%%%%%%%%%%%%%%%%%%%%%%%%%%%%%%%%%%%%%%%%%%%%%%%%%%%%%
\end{document}